\newcommand{\lTeff}{\mbox{$\log T_{\rm eff}$}}
\newcommand{\Mbol}{\mbox{$M_{\rm bol}$}}

\newcommand{\HII}{\mbox{H\hspace{0.2em}{\scriptsize II}}}
\newcommand{\al}{\mbox{$^{26}$\hspace{-0.2em}Al}}
\newcommand{\fe}{\mbox{$^{60}$Fe}}
\newcommand{\ti}{\mbox{$^{44}$Ti}}
\newcommand{\co}{\mbox{$^{56}$Co}}
\newcommand{\yal}{\mbox{$Y_{26}^{\rm O7V}$}}
\newcommand{\yfe}{\mbox{$Y_{60}^{\rm O7V}$}}
\newcommand{\flxal}{\mbox{$S_{1809}$}}
\newcommand{\flxfe}{\mbox{$S_{1137}$}}
\newcommand{\flxff}{\mbox{$S_{53}$}}
\newcommand{\Msol}{\mbox{$M_{\sun}$}}

\newcommand{\Mlow}{\mbox{$M_{\mathrm{low}}$}}
\newcommand{\Mup}{\mbox{$M_{\mathrm{up}}$}}
\newcommand{\gray}{\mbox{$\gamma$-ray}}
\newcommand{\pcmq}{\mbox{cm$^{-2}$}}

\newcommand{\psec}{\mbox{s$^{-1}$}}

\newcommand{\funit}{\mbox{ph \pcmq \psec}}

\newcommand{\flyc}{\mbox{ph \psec}}
\def\MeV{\mbox{Me\hspace{-0.1em}V}}

\def\deg{\ensuremath{^\circ}}

\documentclass{aa}
\usepackage{epsfig}
\begin{document}

\title{Gamma-ray line emission from OB associations and young open 
clusters}

\subtitle{II. The Cygnus region}

\author{J.~Kn\"odlseder\inst{1} 
        \and M.~Cervi\~no\inst{2}  
        \and J.-M.~Le Duigou\inst{1} 
        \and G.~Meynet\inst{3}
        \and D.~Schaerer\inst{4}
        \and P. von Ballmoos\inst{1}}

\offprints{J\"urgen Kn\"odlseder, \email{knodlseder@cesr.fr}}

\institute{Centre d'Etude Spatiale des Rayonnements, CNRS/UPS, B.P.~4346,
	       31028 Toulouse Cedex 4, France
	       \and
	       LAEFF (INTA) Apdo. 50727, Madrid 28080, Spain
	       \and
	       Observatoire de Gen\`eve, CH-1290 Sauverny, Switzerland
	       \and
	       Observatoire Midi-Pyr\'en\'ees, 14, avenue Edouard Belin, 
	       31400 Toulouse, France}

\date{Received / Accepted }

\abstract{
Gamma-ray and microwave observations of the Cygnus region reveal an 
intense signal of 1.809 \MeV\ line emission, attributed to radioactive 
decay of \al, that is closely correlated with 53 GHz free-free emission, 
originating from the ionised interstellar medium.
We modelled both emissions using a multi-wavelength evolutionary 
synthesis code for massive star associations that we applied to the 
known massive star populations in Cygnus.
For all OB associations and young open clusters in the field, we 
determined the population age, distance, and richness as well 
as the uncertainties in all these quantities from published photometric 
and spectroscopic data.
We propagate the population uncertainties in model uncertainties by means 
of a Bayesian method.
The young globular cluster Cyg OB2 turns out to be the dominant \al\ 
nucleosynthesis and ionisation source in Cygnus.
Our model reproduces the ionising luminosity of the Cygnus region very well, 
yet it underestimates \al\ production by about a factor of 2.
We attribute this underestimation to shortcomings of current 
nucleosynthesis models, and suggest the inclusion of stellar rotation 
as possible mechanism to enhance \al\ production.
We also modelled \fe\ nucleosynthesis in the Cygnus region, yet the 
small number of recent supernova events suggests only little \fe\ 
production.
Consequently, a detection of the 1.137 \MeV\ and 1.332 \MeV\ decay lines 
of \fe\ from Cygnus by the upcoming {\em INTEGRAL} observatory is not 
expected.
}

\authorrunning{J. Kn\"odlseder et al.}

\titlerunning{Gamma-ray line emission from OB associations and young open 
clusters}
\maketitle

\section{Introduction}
\label{sec:intro}

OB associations and young open clusters constitute the most prolific 
nucleosynthesis sites in our Galaxy.
The combined activity of stellar winds and core-collapse supernovae 
ejects significant amounts of freshly synthesised nuclei into the 
interstellar medium.
Radioactive isotopes, such as \al\ or \fe, that have been 
co-produced in such events may eventually be observed by gamma-ray 
instruments through their characteristic decay-line signatures.
Indeed, galactic 1.809 \MeV\ gamma-ray line emission attributed to 
the radioactive decay of \al\ has been observed by numerous gamma-ray 
telescopes (see Prantzos \& Diehl 1996\nocite{prantzos96} for a review).
In particular, the COMPTEL telescope provided the first image 
of the Galaxy in the light of this isotope, showing an asymmetric ridge 
of diffuse emission along the galactic plane with a prominent localised 
emission enhancement in the Cygnus region (Diehl et al.~1995\nocite{diehl95}).
The Cygnus emission has been interpreted as the result of \al\ ejection in 
Wolf-Rayet winds and during core collapse supernova explosions from a nearby 
(1-2 kpc) massive star population which probably is part of the local spiral 
arm structure (del Rio et al.~1996\nocite{delRio96}).

Gamma-ray line emission is not the only tracer of this activity.
\cite{knoedl99a} demonstrated that the galactic 1.809 \MeV\ emission is 
closely correlated to galactic free-free emission as observed in the 
microwave domain.
The free-free emission mainly results from the ionisation of the 
interstellar medium by the UV flux of O stars, hence it traces the 
massive star population.
The observed correlation is indeed one of the strongest arguments in 
favour of prolific \al\ production by massive stars. 
In general, a wealth of distinct massive star populations of different 
ages, sizes, or metallicities contribute to the emission along a line 
of sight through the Galaxy, and the observed correlation allows only 
conclusions about the average properties of the contributing 
populations.
In that way, the observations suggest that the equivalent O7V star \al\ 
yield, defined as the average amount of \al\ ejected per number of 
O7V star (measured by their equivalent ionising production), has a
galaxywide constant value of $\yal = (1.0 \pm 0.3) \times 10^{-4}$ \Msol\
(Kn\"odlseder 1999\nocite{knoedl99}).

It is surprising, however, that the correlation between 1.809 \MeV\ 
and microwave free-free emission also holds for the Cygnus region.
Both gamma-ray and microwave data show a localised emission enhancement 
towards Cygnus, similar in size and relative intensity, resulting in 
an equivalent O7V star \al\ yield of 
$(1.1 \pm 0.3) \times 10^{-4}$ \Msol\ (see Sect. \ref{sec:results}).
Within the uncertainties this yield is identical to the galactic value.
In contrast to the Galaxy, however, only few massive star associations 
contribute to the observed emission in Cygnus, and it is not expected 
that they have the same properties as the Galaxy as a whole.
In particular, the galactic metallicity gradient leads to an average 
galactic abundance that is supersolar, and indeed only a supersolar abundance is 
able to reconcile theoretical \al\ yields with the observed galactic \al\ 
mass (Kn\"odlseder 1999\nocite{knoedl99}).
In contrast, massive star populations in Cygnus show slightly subsolar 
abundances (e.g.~Daflon et al.~2001\nocite{daflon01}) and since 
\al\ yields are believed to depend on metallicity (e.g.~Prantzos \& 
Diehl 1996\nocite{prantzos96}) the nucleosynthetic properties of the 
Cygnus region should deviate from those of the average Galaxy.

To understand the observations, we present in this paper a bottom-up 
model of the Cygnus region where we aim to explain the gamma-ray and 
microwave data from the underlying stellar populations.
For this purpose we developed a multi-wavelength evolutionary synthesis 
model that we presented in paper I of this series 
(\nocite{cervino00}Cervi\~no et al.~2000).
We put considerable effort into the characterisation of the massive star 
populations in Cygnus with particular emphasis on the involved 
uncertainties (distance and age uncertainty; coeval or continuous star 
formation).
We incorporate these uncertainties into our model by means of a 
Bayesian method and determine confidence intervals for all quantities 
to assess the predictive power of our approach.
Despite the resulting uncertainties, we will demonstrate that the 
gamma-ray observations provide important clues on nucleosynthesis 
physics in massive stars.
In particular we will demonstrate the shortcomings of current 
theoretical nucleosynthesis models in explaining \al\ production and 
discuss possible modifications that may improve the models.

\section{Evolutionary synthesis model and analysis method}
\label{sec:model}

The evolutionary synthesis model we employed in this work is described in 
detail in the first paper of this series by \cite{cervino00}.
In summary, the evolution of each individual star in a stellar population 
is followed using Geneva evolutionary tracks with enhanced 
mass-loss rates (\nocite{meynet94}Meynet et al.~1994).
Stellar Lyman continuum (Lyc) luminosities are predicted using the CoStar
atmosphere models of \cite{schaerer97}, supplemented by the 
\cite{schmutz92} atmospheres for the Wolf-Rayet phase.
At the end of stellar evolution, stars initially more massive than 
$M_{\rm WR}=25$ \Msol\ are exploded as Type Ib supernovae, while stars 
of initial mass within $8\Msol$ and $M_{\rm WR}$ are assumed to 
explode as Type II SNe.
Nucleosynthesis yields have been taken from \cite{meynet97} for the 
pre-supernova evolution, from \cite{woosley95} for Type II 
and from \cite{wlw95} for Type Ib supernova explosions.
Note that Type II SN yields have only been published for stars without 
mass loss and Type Ib yields have only been calculated for pure Helium 
stars.
In order to obtain consistent nucleosynthesis yields for Type II 
supernovae we followed the suggestion of \cite{maeder92} and linked 
the explosive nucleosynthesis models of \cite{woosley95} to the Geneva 
tracks via the carbon-oxygen core mass at the beginning of carbon burning.
For Type Ib SN we used the helium-core mass at the beginning of helium core 
burning for the link.

In order to predict the present gamma-ray line emission from a massive 
star association we have to estimate the production of radioactive 
isotopes in the past.
We do this by determining the actual number of stars $N_{\ast}$ 
within a given initial mass interval $[\Mlow,\Mup]$ that 
is not affected by incompleteness at the lower end and evolutionary 
effects at the upper end.
We then extrapolate this population back to the past using a Salpeter 
initial mass function (IMF) of slope $\Gamma=-1.35$ that is normalised 
to the number of stars we observe today.
In practice, we generate an initial stellar population by randomly 
sampling the IMF until the number of stars falling in the mass interval 
$[\Mlow,\Mup]$ amounts to $N_{\ast}$. 
We then use the evolutionary synthesis code to follow the time evolution 
of the \al\ and \fe\ yields, the Lyc luminosity, and the distribution 
of spectral types within the population.

The stellar systems we study in this work only comprise a few 
hundred to a few thousand stars, making the high-mass end of the 
population generally sparsely sampled.
However, massive stars provide a considerable fraction of the 
association nucleosynthesis and ionising power, and the early 
evolution of the system will depend rather sensitively on the actual 
choice of stellar masses.
In particular, different Monte Carlo samples of the same initial stellar 
population may lead to quite different evolutions of the association 
observables, leading to considerable uncertainties in our model 
predictions.
We decided to include these uncertainties in our analysis by means of 
a Bayesian method.
Instead of investigating the time evolution of a quantity $X$ for a 
specific population, we base our analysis on the probability density 
function (PDF) $p(X|t)$ which quantifies the probability for $X$ to be 
equal to a value $x$ at the time $t$.
We approximate this function for a given association by repeating the 
evolutionary synthesis calculations for independently sampled stellar 
populations, from which we obtain the frequency of observing the value 
$x$ at time $t$.
In this way, the uncertainty arising from the extrapolation to the 
past, which manifests in an uncertainty about the initial massive star 
population, is reflected in the width of the PDF.
Examples of PDFs obtained for massive star associations in the Cygnus 
regions are given in \cite{cervino00}.

To predict the today value of the quantity $X$, we marginalise over the 
(uncertain) age $t$ of the association using
\begin{equation}
 p(X) = \int_{0}^{\infty} p(t) p(X|t) dt ,
 \label{eq:age}
\end{equation}
where the prior probability distribution $p(t)$ quantifies the age 
uncertainty.
We tried different choices for the prior distributions, such as 
Gaussians or bounded uniform functions, but we found that the precise 
form of the prior has little impact on the result as long as it 
quantifies our knowledge about the age uncertainty (see Sect. \ref{sec:prior}).
To estimate fluxes $S_{i}$ from luminosities $L_{i}$ for a given 
association $i$, we further marginalise over the distance $s$ of the 
association using
\begin{equation}
 p_{i}(S_{i}) = \int_{0}^{\infty} p(s) p(L_{i}|s) ds .
 \label{eq:distance}
\end{equation}
where the prior $p(s)$ quantifies the distance uncertainty, and
$S_{i} \propto s^{-2}$.
Integral fluxes from the Cygnus region were obtained by adding the 
contributions from individual OB associations and young open clusters 
using successive marginalisation, i.e.
\begin{equation}
 p(S) = \int_{0}^{\infty} p_{1}(S_{1}) p_{2}(S-S_{1}) d S_{1} ,
 \label{eq:total}
\end{equation}
where $p_{1}$ and $p_{2}$ are the flux PDFs for association $1$ and 
$2$, respectively, and $S = S_{1} + S_{2}$.
Finally, we derive mean (or median) flux values and 63.8 \% confidence 
regions from these PDFs which we then compare to observations.

\section{Massive star populations in Cygnus}
\label{sec:data}

The starting point of our study of massive star populations in Cygnus 
is the WEBDA database of open clusters of \cite{mermilliod98}.
From this database we established a list of open clusters within the 
limits $60\deg < l < 110\deg$ and $|b| < 15\deg$, a region which 
largely encloses the 1.809 \MeV\ and microwave free-free emission 
features.
We preselected young clusters from this list by requiring either an 
earliest spectral type of B3V or earlier (corresponding to a star of 
$\sim8$ \Msol\ initial mass), or a cluster age of less than 
$5 \times 10^7$ yr.
Older clusters do not house any potential core collapse progenitor
anymore, and their nucleosynthesis activity is therefore negligible.
The lack of stars more massive than $\sim10$ \Msol\ also reduces their 
ionising flux to an imperceptible level.
Of course, not for all known open clusters an age estimate is 
available, but those clusters are then often rather distant and/or 
do not contain luminous members.
Consequently, they do not contribute significantly to the emissions 
that we aim to study in this work.

For the OB associations in the Cygnus region we gathered the relevant 
information from the literature.
Our main resources were the compilations of \cite{humphreys78} and 
\cite{garmany92}, complemented with data from the SIMBAD database and 
some recent investigations (see the appendix for a detailed description 
of our database).

The association parameters relevant to our study are the age $\tau$, 
the distance $d$, the richness, expressed as the number of 
stars $N_{\ast}$ in a specific initial mass interval $[\Mlow, \Mup]$, 
and the slope $\Gamma$ of the IMF
(in the following we deliberately choose the term association when we 
talk both about OB associations and young open clusters).
Literature values for these parameters are only available for some of 
the associations, and have generally been derived by different methods.
Also the choice of the stellar evolutionary tracks and luminosity 
calibrations impacts the results, and may lead to differences among 
the derived parameters.
Additionally, the uncertainty of the association parameters has rarely 
been determined, and if so, also not in a consistent manner.
For these reasons we decided to re-determine the relevant association 
parameters and their uncertainties from the basic stellar data in our 
database.
In the following we describe our method and present the resulting 
parameters for all young stellar associations in Cygnus.
The results of the analysis are summarised in Tables \ref{tab:distance} 
and \ref{tab:obdata}.
Details for each association are given in the appendix.

\subsection{Reddening}
\label{sec:reddening}

As a first step we determine the reddening for each association from 
spectroscopic and photometric information.
For each star with spectral classification we used the spectral type to 
intrinsic colour calibration of \cite{fitzgerald70} complemented by information 
from \cite{schmidtkaler82} for the earliest spectral types to determine 
their colour excess $E(B-V)=(B-V)-(B-V)_{0}$ and $E(U-B)=(U-B)-(U-B)_{0}$.
From these we calculate the slope $q_{r} = E(U-B)/E(B-V)$ of the reddening curve 
for each association by averaging the slopes of all member stars with 
known spectral type.
Outliers were iteratively removed from the average until all 
remaining stars lie within 2 rms of the average.

The slopes typically vary from $0.67$ to $0.84$ with a few exceptions
with poorly defined values due to the proximity of the association
(Cyg OB7, Lac OB1, Roslund 5).
We find a trend towards steeper slopes (typically around $q_{r} = 
0.80$) for the more reddened associations, in particular in the area
of Cyg OB1, OB2 and OB9, i.e.~the central part of the Cygnus X region.
This confirms findings by other authors 
(e.g.~Massey \& Thompson 1991\nocite{massey91}) of an anomalous reddening 
in this area.
For the couple of open clusters for which no spectral data are 
available, the canonical value of $q_{r} = 0.72$ has been adopted.
In view of the dispersion in the value of $q_{r}$ in the Cygnus area this 
choice is certainly somewhat arbitrary, but the impact of the exact 
value on the resulting cluster parameters is rather small, leading to 
a negligible additional uncertainty in our study.

\subsection{Distance}
\label{sec:distance}

\begin{table*}[th]
  \footnotesize
  \caption{\label{tab:distance}
  Reddening and distance moduli for OB associations and young open clusters in 
  the field $60\deg<l<110\deg$, $-15\deg<b<15\deg$. 
  Columns 5 and 6 specify the selection criteria that were applied to remove 
  field or background stars from the datasets.
  Column 7 indicates the possible physical relation of an open 
  cluster to an OB association as suggested by our analysis (see 
  details in the appendix).
  }
  \begin{flushleft}
    \begin{tabular}{llrcccl}
    \hline
    \noalign{\smallskip}
    Name &
    \multicolumn{1}{c}{$q_{r}$} &
    \multicolumn{1}{c}{$DM$} & 
    $E(B-V)$ & 
    \multicolumn{1}{c}{$E(B-V)$ range} & 
    \multicolumn{1}{c}{$\Delta DM$} &
    \multicolumn{1}{l}{Association} \\
    \noalign{\smallskip}
    \hline
    \noalign{\smallskip}
    Cep OB1 & $0.75\pm0.03$ & $12.8\pm0.4$ & $0.60\pm0.16$ & - & - & \\
    Cep OB2 & $0.79\pm0.13$ &  $9.1\pm0.3$ & $0.49\pm0.14$ & - & - & \\
    Cyg OB1 & $0.80\pm0.04$ & $11.4\pm0.4$ & $0.74\pm0.15$ & - & 0.8 & \\    
    Cyg OB2 & $0.79\pm0.02$ & $11.0\pm0.5$ & $1.84\pm0.29$ & $1.2-5.0$ & 0.8 & \\
    Cyg OB3 & $0.73\pm0.04$ & $11.7\pm0.7$ & $0.49\pm0.12$ & - & - & \\
    Cyg OB7 & $0.66\pm0.20$ &  $9.6\pm0.2$ & $0.36\pm0.23$ & - & - & \\
    Cyg OB8 & $0.77\pm0.04$ & $11.9\pm0.5$ & $0.97\pm0.33$ & - & - & \\
    Cyg OB9 & $0.81\pm0.06$ & $10.5\pm0.4$ & $1.12\pm0.26$ & - & - & \\
    Lac OB1 & $0.58\pm0.31$ &  $9.0\pm0.5$ & $0.12\pm0.03$ & - & - & \\
    Vul OB1 & $0.69\pm0.06$ & $11.9\pm0.2$ & $0.92\pm0.07$ & $0.8-1.1$ & 0.8 & \\
    \noalign{\smallskip}
    \hline
    \noalign{\smallskip}
    Berkeley 86  & $0.79\pm0.05$ & $11.1\pm0.3$ & $0.96\pm0.12$ & -          & 0.6 & Cyg OB1 \\
    Berkeley 87  & $0.84\pm0.03$ & $11.4\pm1.0$ & $1.63\pm0.13$ & -          & -   & Cyg OB1 \\
    Berkeley 94  & $0.72$        & $13.6\pm0.1$ & $0.68\pm0.06$ & -          & -   & \\
    Berkeley 96  & $0.72$        & $13.6\pm0.1$ & $0.65\pm0.04$ & -          & -   & \\
    Biurakan 2   & $0.72$        & $11.8\pm0.4$ & $0.47\pm0.08$ & $0.2-0.7$  & 0.8 & Cyg OB3 (?) \\
    IC 4996      & $0.78\pm0.12$ & $11.1\pm0.3$ & $0.68\pm0.06$ & $0.5-0.8$  & 1.0 & Cyg OB1 \\
    IC 5146      & $0.72\pm0.18$ & $10.3\pm1.0$ & $0.64\pm0.34$ & $0.3-1.2$  & 1.2 & \\
    NGC 6823     & $0.70\pm0.08$ & $12.1\pm0.4$ & $0.86\pm0.14$ & $0.5-1.2$  & 0.8 & Vul OB1 \\
    NGC 6871     & $0.81\pm0.09$ & $11.9\pm0.4$ & $0.45\pm0.02$ & $0.4-0.6$  & 0.6 & Cyg OB3 \\
    NGC 6883     & $0.72$        & $11.3\pm0.7$ & $0.39\pm0.04$ & -          & -   & Cyg OB3 (?) \\
    NGC 6910     & $0.80\pm0.06$ & $11.3\pm0.5$ & $1.12\pm0.09$ & $0.9-1.25$ & 0.8 & \\
    NGC 6913     & $0.80\pm0.15$ & $11.3\pm1.0$ & $0.91\pm0.14$ & $0.6-1.5$  & -   & Cyg OB1 \\
    NGC 7067     & $0.72$        & $14.2\pm0.1$ & $0.90\pm0.10$ & -          & -   & \\
    NGC 7128     & $0.72$        & $13.2\pm0.4$ & $1.05\pm0.06$ & $0.7-1.3$  & 0.8 & \\
    NGC 7160     & $0.73\pm0.16$ & $10.0\pm0.4$ & $0.35\pm0.04$ & $0.2-0.8$  & 0.6 & \\
    NGC 7235     & $0.67\pm0.06$ & $12.8\pm0.4$ & $0.92\pm0.04$ & $0.7-1.1$  & 0.8 & Cep OB1 \\
    NGC 7261     & $0.72$        & $12.8\pm0.4$ & $1.04\pm0.10$ & $0.8-1.4$  & 0.8 & Cep OB1 \\
    NGC 7380     & $0.78\pm0.06$ & $12.8\pm0.2$ & $0.60\pm0.04$ & $0.5-0.9$  & 0.8 & Cep OB1 \\
    Roslund 4    & $0.72$        & $12.3\pm0.6$ & $1.05\pm0.15$ & -          & -   & \\
    Roslund 5    & $1.15\pm0.16$ &  $8.6\pm0.3$ & $0.10\pm0.04$ & -          & -   & \\
    Ruprecht 175 & $0.63\pm0.18$ & $11.3\pm0.1$ & $0.26\pm0.03$ & -          & -   & \\
    Trumpler 37  & $0.71\pm0.07$ & $10.2\pm0.4$ & $0.54\pm0.08$ & $0.2-1.0$  & 0.8 & \\
    \noalign{\smallskip}
    \hline
    \end{tabular}
  \end{flushleft}
\end{table*}

As a second step the distance of each association is estimated.
For associations with spectroscopic information we employed the 
method of spectroscopic parallax.
For each MK classified star we compute $DM=V-A_{V}-M_{V}$, where 
$A_{V}=R_{V} \times E(B-V)$ is the visible extinction.
The association distance is then calculated by averaging $DM$ 
where again 2 rms outliers were removed iteratively.
The absolute visual magnitudes $M_{V}$ have been extracted from 
calibrations of \cite{vacca96} for stars earlier than B1, 
complemented by data from \cite{humphreys84} and \cite{schmidtkaler82}
for later spectral types.
For the ratio $R_{V}$ of absolute-to-selective absorption we apply the 
canonical value of $3.1$, although we recognise that deviations from 
this value are eventually observed in star forming regions 
(\nocite{mathis90}Mathis 1990).
For Cyg OB2, for example, \cite{massey91} found $R_{V}=3.0$, and 
indeed a variable extinction analysis of our data suggests a similar 
value.
This deviation results in a distance modulus error of $0.2$ mag for 
this association, inferior to the statistical and systematical uncertainty 
of $0.5$ mag.
Cyg OB2 is certainly an extreme case due to the large reddening of 
the association, and we believe that variations in $R_{V}$ should not 
dramatically alter our results.
Anyways, we have little choice since generally the available data do 
not allow to perform a reliable variable extinction analysis for the 
associations, and we fear that a poorly determined value of $R_{V}$ 
introduces a larger error than a rather solid mean value of $3.1$ 
that may not precisely apply to all cases.

For some of the open clusters for which not enough spectral type
information was available, we estimated the distance from the hot 
cluster stars using the reddening-free parameter 
$Q=(U-B)-q_{r} \times (B-V)$, where $q_{r}$ is the mean cluster reddening 
slope as given in Table \ref{tab:distance}.
For $Q<-0.4$, corresponding to spectral types earlier than B5 or so,
there exist unique relations between intrinsic colour $(B-V)_{0}$, absolute 
visual magnitude $M_{V}$, and $Q$ for each luminosity class.
We determined these relations by fitting polynomial functions to 
calibration tables that we compiled from \cite{fitzgerald70}, 
\cite{schmidtkaler82}, \cite{humphreys84}, and \cite{vacca96}.
To estimate the luminosity class we apply an iterative approach.
For each star with $Q<-0.4$ we assume as initial estimate a luminosity 
class of V (this is certainly a reasonable assumption for young unevolved 
clusters).
Based on this estimate we use the calibration relations to determine 
$(B-V)_{0}$ and $M_{V}$ from $Q$.
From these quantities we derive $E(B-V)$, $A_{V}$, and $DM$,
leading to a first cluster distance estimate by averaging over $DM$.
Cluster stars which we misclassified in luminosity class will clearly 
show up as outliers in $DM$ with respect to the average.
Hence, for stars with $DM$ more than 1 rms above the average we reduce 
the luminosity class (in the sense V $\to$ I) while for stars with
$DM$ more than 1 rms below the average we increase the luminosity class 
(in the sense I $\to$ V) for the next iteration.
We then repeat the entire procedure for the updated luminosity classes.
For all clusters in our sample, this scheme led to convergence in 
$DM$ after $2-3$ iterations.
We tested the validity of our procedure by applying the method to clusters for 
which $DM$ is known from spectroscopic parallaxes and found satisfactory 
agreement with the photometric results.

\subsection{Field star separation}
\label{sec:fieldstar}

For Cyg OB2 and many of the clusters, field star contamination is a 
serious problem in our database.
Field stars may be recognised by particularly low or high $E(B-V)$ 
values with respect to bona fide association members 
(e.g.~\nocite{massey95}Massey et al.~1995).
For contaminated associations, we therefore excluded those stars whose 
inferred colour excess is outside the range spanned by member stars 
with MK classification.
Most clusters show a clear upper main sequence which allows for an 
unambiguous definition of stellar reddenings of cluster members.
For the cases where we applied such a selection, the $E(B-V)$ range 
of cluster members is specified in column 5 of Table \ref{tab:distance}.

To perform such a selection, $E(B-V)$ has been determined for each star
from either spectroscopic or photometric data.
In the latter case, the iterative procedure described above has been 
used for hot stars with $Q<-0.4$.
For $Q\ge-0.4$, the relation between $(B-V)_{0}$ and $Q$ becomes 
ambiguous, corresponding to the well known `knee' in the colour-colour 
diagram.
Nevertheless, from the intersection of the reddening lines with the 
intrinsic colour-colour tracks we can determine all possible $E(B-V)$ 
values (3 at maximum) that are consistent with the observed colours, and
we exclude all stars that have no solution that is compatible with the 
specified colour excess interval.

To remove remaining field or background stars we employed for most
clusters another selection criterion based on the distance 
moduli of the stars.
From bona fide members with MK classification we define a distance moduli 
interval $\Delta DM$ and exclude all stars for which 
$|DM - \langle DM \rangle| > \Delta DM$, where $\langle DM \rangle$ is 
the distance modulus of the association.
Distance moduli for stars with MK classification or hot ($Q<-0.4$) stars 
were determined from spectroscopic and photometric parallaxes, 
respectively.
For the remaining stars, distance moduli were again estimated from 
the intersection of the reddening lines with the intrinsic colour-colour 
tracks.

\subsection{Transformation to \lTeff\ and \Mbol}
\label{sec:trafo}

\begin{figure}[tb]
  \includegraphics[width=8.8cm]{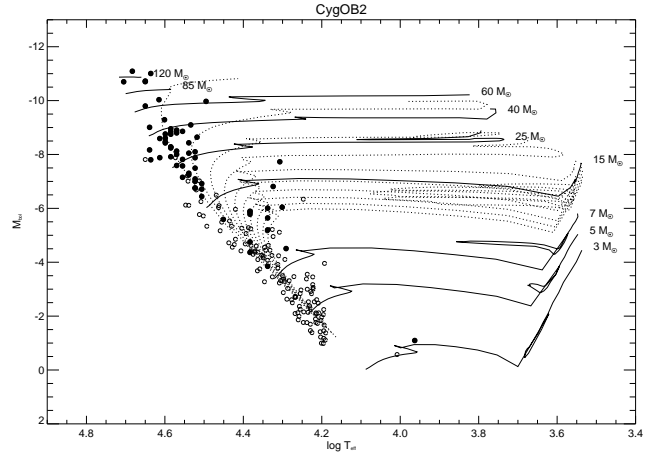}
  \caption{H-R diagram of Cyg OB2. The filled circles are stars for which 
           spectral classification is available, the open circles are stars 
           for which we have only photometry. Solid lines are the stellar 
           tracks of \cite{meynet94}, dotted lines are the corresponding 
           isochrones for intervals of 2 Myr, starting from $\tau=2$ Myr.}
   \label{fig:CygOB2}
\end{figure}

Our next step consists of constructing H-R diagrams (HRDs) for all 
associations.
For this purpose we transform the spectroscopic and photometric 
information for all member stars to \lTeff\ and \Mbol.
If a star has a MK classification, we used the spectral type to determine 
both the effective temperature and bolometric correction $B.C.$ using the 
calibration of \cite{vacca96}, complemented by data from \cite{humphreys84} and 
\cite{schmidtkaler82}.
From the absolute visual magnitude and the bolometric correction, the bolometric 
luminosity is then derived using $\Mbol = M_{V} + B.C.$

\begin{table*}[th]
  \footnotesize
  \caption{\label{tab:obdata}
  Population parameters for the associations of Table 
  \ref{tab:distance} (IC 5146 which is older than 50 Myr has been excluded).
  Columns 3 and 4 list the number of O and WR stars within the 
  association, columns 5-7 give the WR subtype distribution.
  The 120 O stars quoted for Cyg OB2 are probably an upper limit (see 
  text) while the belonging of the O star 10 Lac to Lac OB1 is 
  questionable (see appendix).
  The WR identifications are given in the last column.
  The physical reality of the associations marked by $^{\ast}$ is 
  extremely doubtful.
  The last row summarises the stellar statistics for the entire 
  region.
  }
  \begin{flushleft}
    \begin{tabular}{lccccccccl}
    \hline
    \noalign{\smallskip}
    Name & 
    age (Myr) & 
    $N_{O}$ & 
    $N_{WR}$ & 
    $N_{WN}$ & 
    $N_{WC}$ & 
    $N_{WO}$ & 
    $N_{\ast}$ & 
    $[\Mlow, \Mup]$ & 
    \multicolumn{1}{c}{WR} \\
    \noalign{\smallskip}
    \hline
    \noalign{\smallskip}
    Cep OB1a          &  $2-5$ &  18 & 4 & 3 & 1 & - &  16 &  $[20,40]$ & 152,153,154,155 \\
    Cep OB1b          & $9-18$ &   - & - & - & - & - &  19 &  $[ 9,15]$ & \\
    Cep OB2           &$10-16$ &   - & - & - & - & - &   9 &  $[ 7,12]$ & \\
    Cyg OB1           &  $2-6$ &  11 & 4 & 3 & 1 & - &  23 &  $[15,40]$ & 136,137,138,141 \\    
    Cyg OB2           &  $1-4$ & (120) & 3 & 1 & 2 & - & 120 & $[20,120]$ & 144,145,146 \\    
    Cyg OB3           &  $2-5$ &  12 & 2 & 1 & 1 & - &  14 &  $[25,60]$ & 134,135 \\
    Cyg OB7           &  $2-5$ &   2 & - & - & - & - &  10 &   $[7,25]$ & \\
    Cyg OB8$^{\ast}$  & $1-14$ &   6 & - & - & - & - &   8 &  $[20,40]$ & \\
    Cyg OB9           &  $2-5$ &   8 & - & - & - & - &   6 &  $[20,40]$ & \\
    Lac OB1           &$12-15$ & (1) & - & - & - & - &  14 &   $[7,12]$ & \\
    Vul OB1$^{\ast}$  &  $2-6$ &   3 & - & - & - & - &   5 &  $[15,40]$ & \\
    \noalign{\smallskip}
    \hline
    \noalign{\smallskip}
    Berkeley 86  &   $3-5$ & 2 & 1 & 1 & - & - & 11 &  $[7,25]$ & 139 \\
    Berkeley 87  &   $3-6$ & 2 & 1 & - & - & 1 & 24 &  $[7,25]$ & 142 \\
    Berkeley 94  &   $3-5$ & 1 & - & - & - & - &  4 &  $[7,15]$ & \\
    Berkeley 96  &   $4-7$ & 1 & - & - & - & - &  3 &  $[7,15]$ & \\
    Biurakan 2   & $24-30$ & - & - & - & - & - & 12 &  $[4, 9]$ & \\
    IC 4996      &   $4-7$ & 2 & - & - & - & - &  6 &  $[7,25]$ & \\
    NGC 6823     &   $1-5$ & 4 & - & - & - & - & 21 &  $[7,20]$ & \\
    NGC 6871     &   $5-6$ & - & 1 & 1 & - & - & 13 &  $[7,25]$ & 133 \\
    NGC 6883     & $14-16$ & - & - & - & - & - &  2 & $[12,15]$ & \\
    NGC 6910     &   $4-5$ & 2 & - & - & - & - &  7 &  $[7,25]$ & \\    
    NGC 6913     &   $2-5$ & 3 & - & - & - & - & 13 &  $[7,40]$ & \\    
    NGC 7067     & $14-18$ & - & - & - & - & - & 27 &   $[5,9]$ & \\        
    NGC 7128     & $26-30$ & - & - & - & - & - & 27 &   $[4,7]$ & \\    
    NGC 7160     &  $8-14$ & - & - & - & - & - &  4 &   $[5,7]$ & \\    
    NGC 7235     &   $4-5$ & - & - & - & - & - & 18 &  $[7,15]$ & \\    
    NGC 7261     & $12-20$ & - & - & - & - & - & 17 &  $[5,12]$ & \\    
    NGC 7380     &   $4-5$ & 1 & - & - & - & - &  8 &  $[7,20]$ & \\    
    Roslund 4    &  $5-15$ & - & - & - & - & - &  6 &  $[7,12]$ & \\    
    Roslund 5    & $20-30$ & - & - & - & - & - &  6 &  $[3, 5]$ & \\    
    Ruprecht 175 & $20-30$ & - & - & - & - & - &  3 &  $[3, 5]$ & \\    
    Trumpler 37  &   $3-6$ & 5 & - & - & - & - & 25 &  $[7,25]$ & \\
    \noalign{\smallskip}
    \hline
    \noalign{\smallskip}
    total        & & 204 & 16 & 10 & 5 & 1 & & & \\
    \noalign{\smallskip}
    \hline
    \end{tabular}
  \end{flushleft}
\end{table*}

For stars without MK classification, we again distinguish between hot ($Q<-0.4$) 
and cold ($Q\ge-0.4$) stars.
For hot stars we determined \lTeff\ and $B.C.$ from $Q$, while for 
cold stars we derived both parameters from the intrinsic colour $(B-V)_{0}$.
Using the mean association distance $DM$ and extinction $A_{V}$ we 
assign a bolometric luminosity to each star using
$\Mbol = V - A_{V} - DM + B.C.$
We estimate the intrinsic colour $(B-V)_{0} = (B-V) - E(B-V)$ for each star by 
assuming a reddening identical to the mean reddening of the association.
Our analysis is based on polynomial functions that we fitted to the calibration 
tables of intrinsic stellar parameters.
In general, the relations depend of the luminosity class which we 
estimate for each star from the absolute visual magnitude
$M_{V}=V - A_{V} - DM$ (unless it has already been estimated by the iterative 
procedure described in \S\ref{sec:distance}).
Here we make use of the rapid variation of $M_{V}$ with luminosity 
class, and we search the calibration tables for the luminosity class 
that comes closets to the pair $M_{V}$ and $(B-V)_{0}$ of stellar 
parameters.
Although this method is certainly not very accurate, the impact of a 
misclassification in luminosity class is only moderate.
Typically, \lTeff\ varies about $0.02-0.1$ dex and $B.C.$ about 
$0.3-0.7$ mag between subsequent luminosity classes.

\subsection{Age determination}
\label{sec:age}

From the H-R diagrams we estimated the age of the associations.
For this purpose we compared the location of the association stars in 
the HRDs to the theoretical evolutionary tracks of \cite{meynet94} for 
solar metallicity ($Z=0.02$).
The same tracks are used for the evolutionary synthesis calculations, 
hence in this respect our models are consistent with the population 
input data.
As an example we show the HRD for Cyg OB2 in Fig.~\ref{fig:CygOB2} which 
may be compared to Fig.~16 of \cite{massey91} who made a similar analysis.
From the upper part of the HRD (masses $\ga 20$ \Msol) we infer the ages 
of the massive stars from the \cite{meynet94} isochrones that we 
superimposed on the diagrams.
For Cyg OB2, for example, this method leads us to an age estimate of 
$1-4$ Myr, comparable to the finding of \cite{herrero99}.

The results for all associations are summarised in Table \ref{tab:obdata}.
Instead of determining a best fitting age, we always tried to specify an age 
range that reflects the dispersion of the upper main sequence.
The dispersion comes partly from the systematic limitation of our 
method (knowledge of stellar calibrations, neglection of stellar 
rotation and binarity, measurement or classification errors), but 
could also reflect a period of continuous star formation.
In particular, some associations like Cep OB1, Cyg OB8 or Roslund 4 
show a considerable scatter in the upper HRD which could even reflect 
several epochs of massive star formation.
For Cep OB1 we decided to split the population into two subgroups of 
different age since the data are difficult to reconcile with a single 
star formation event.
Yet, a large apparent age spread could also signal that the 
association presents not a physically associated group but rather a 
chance projection of massive stars in the sky.
For this reason we excluded Cyg OB4 from our study since 
{\em Hipparcos} data did not confirm the reality of this stellar group
(de Zeeuw et al.~1999\nocite{dezeeuw99}).
Other associations such as Cyg OB8 or Vul OB1 are also highly 
doubtful, yet it will turn out that they provide only a negligible 
contribution to the overall luminosities in the Cygnus region, hence 
we do not introduce an important uncertainty by including them into 
our study.

\subsection{Association richness and IMF}
\label{sec:masses}

\begin{table}[t!]
  \footnotesize
  \caption{\label{tab:imf}
  IMF slopes $\Gamma$ for two OB associations and some open clusters.
  The third column indicates the mass range over which the slope has
  been determined.
  }
  \begin{flushleft}
    \begin{tabular}{lcc}
    \hline
    \noalign{\smallskip}
    Name & $\Gamma$ & mass range (\Msol) \\
    \noalign{\smallskip}
    \hline
    \noalign{\smallskip}
    Cyg OB1     & $-1.0\pm0.4$ &  $15- 85$ \\
    Cyg OB2     & $-1.1\pm0.3$ &  $15-120$ \\
    \noalign{\smallskip}
    \hline
    \noalign{\smallskip}
    Berkeley 86 & $-1.2\pm0.5$ & $7-40$ \\
    Berkeley 87 & $-1.2\pm0.4$ & $7-60$ \\
    IC 4996     & $-1.6\pm0.5$ & $5-40$ \\
    NGC 6823    & $-1.3\pm0.4$ & $7-60$ \\
    NGC 6871    & $-1.0\pm0.5$ & $7-40$ \\
    NGC 6910    & $-0.9\pm0.4$ & $5-40$ \\    
    NGC 6913    & $-0.6\pm0.5$ & $7-60$ \\    
    NGC 7235    & $-2.6\pm0.6$ & $7-40$ \\    
    NGC 7380    & $-0.9\pm0.5$ & $7-40$ \\    
    Trumpler 37 & $-1.1\pm0.4$ & $7-60$ \\
    \noalign{\smallskip}
    \hline
    \end{tabular}
  \end{flushleft}
\end{table}

The comparison of the H-R diagram with the evolutionary tracks allows 
an estimation of the initial richness of the association.
We determined the richness by counting the number of stars $N_{\ast}$ over 
a sufficiently large initial mass interval $[\Mlow, \Mup]$.
We tried to choose the lower mass limit $\Mlow$ sufficiently high to 
avoid any bias from population incompleteness due to the limiting 
magnitudes of the association surveys.
Conversely, the upper mass limits $\Mup$ were chosen sufficiently 
low so that the evolutionary turn-off has no impact on the richness 
determination.
The results are summarised in columns 8 and 9 of Table \ref{tab:obdata}.

The only exception to this procedure is Cyg OB2 for which we rely on 
the recent mass determination of \cite{knoedl00}, based on 
an analysis of {\em 2MASS} near infrared data.
In fact, the data that are available in the visible waveband for Cyg 
OB2 are heavily affected by interstellar absorption, and their use would 
considerably underestimate the total mass of this association.
In contrast, the {\em 2MASS} data do not allow a precise spectral 
type determination and the quoted number of 120 O stars rather 
reflects the fact that there were initially 120 stars more massive than 
$\sim 20\Msol$ in Cyg OB2, although some may already have 
evolved, having either turned in early B-type or even Wolf-Rayet stars.
The quoted number of 120 O stars should therefore present an 
upper limit.

For some associations our stellar database contains enough objects 
to allow an estimation of the initial mass spectrum.
We fitted these spectra by power-law initial mass functions and 
summarise their slopes $\Gamma$ in Table \ref{tab:imf}.
We also quote the mass intervals that have been used for the fitting 
which again were chosen to minimise any bias due to evolved stars, 
population incompleteness, and field star contamination.
Yet we recognise that the results are still affected by systematic 
uncertainties, as demonstrated by the steep slope of 
$\Gamma=-2.6\pm0.6$ for NGC 7235 which shows a heavy field star 
contamination (Massey et al.~1995\nocite{massey95}).
Formally, we derive a weighted mean IMF slope of $\Gamma=-1.2\pm0.1$ 
which is close to the canonical Salpeter value of $-1.35$ and 
compatible with the value of $\Gamma=-1.12\pm0.08$ that has been 
determined by \cite{massey95} for 12 galactic OB associations and open 
clusters.
To illustrate the systematic uncertainties in IMF slope 
determinations, we compare our result for Cyg OB2 ($-1.1\pm0.3$) to 
the value of $-1.0\pm0.1$ determined by \cite{massey91} from the same 
stellar data, and the recent analysis of \cite{knoedl00} of 
{\em 2MASS} near infrared data that suggests $\Gamma=-1.6\pm0.1$.
In view of these uncertainties, we believe that the assumption of a 
uniform canonical Salpeter law for our evolutionary synthesis modelling 
of all associations is a reasonable approximation that is compatible 
with the observational data.

\begin{figure*}[tb]
  \includegraphics[width=8.9cm]{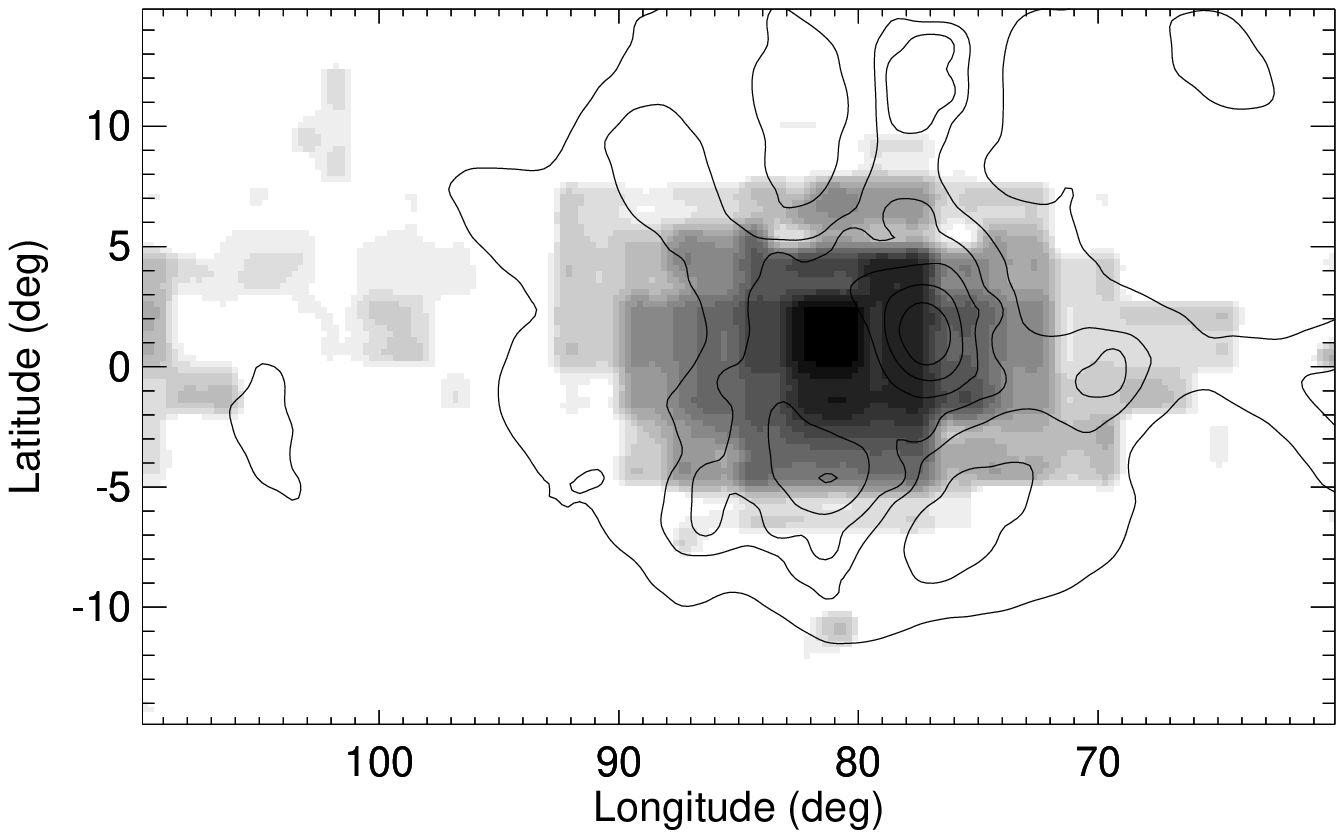} \hfill
  \includegraphics[width=8.9cm]{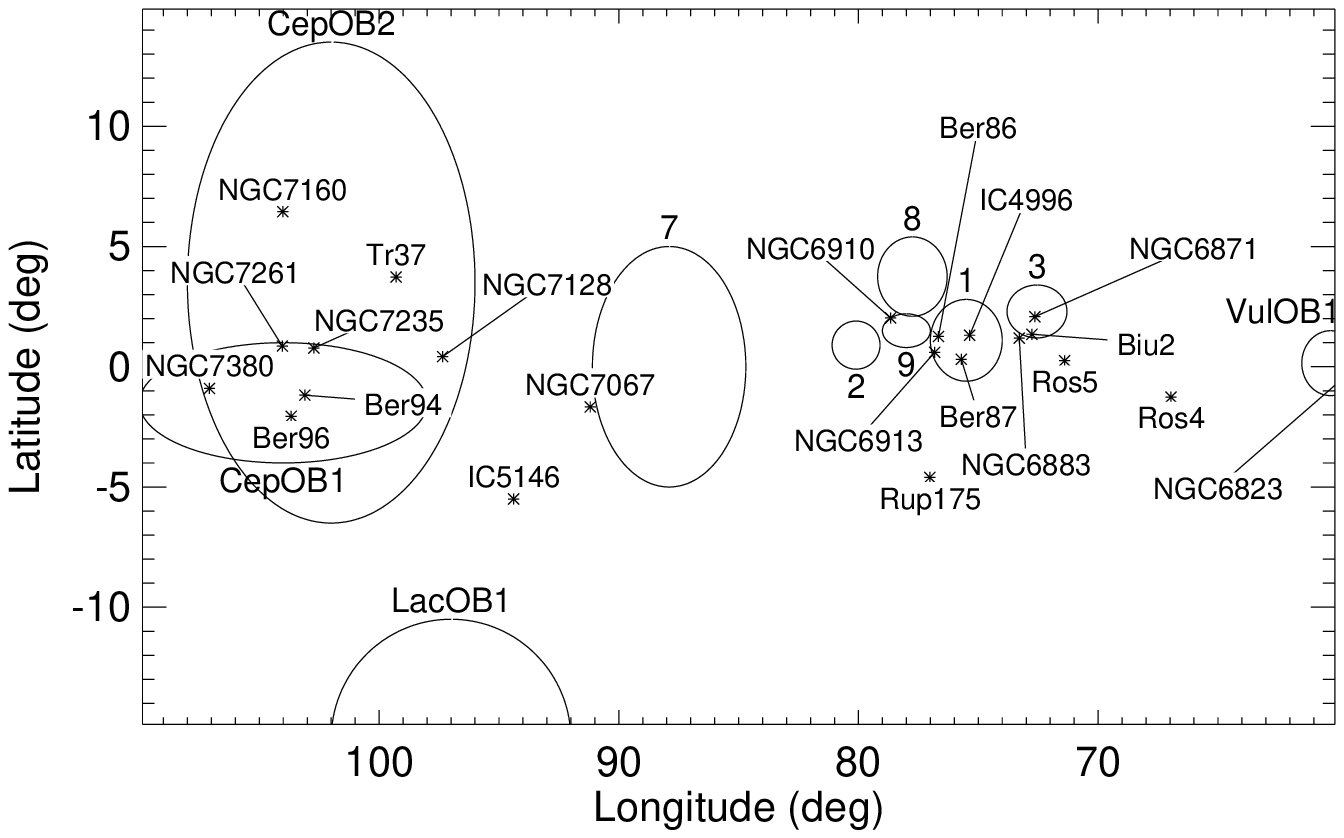}
  \caption{{\em Left:} Contour map of 1.809 \MeV\ gamma-ray line emission in 
           the Cygnus region (Pl\"uschke 2001) superimposed 
           onto a greyscale image of microwave 53 GHz free-free emission 
           (Bennett et al.~1992).
           The prominent emission feature extending from $l\sim70\deg$ to 
           $90\deg$ and $b\sim \pm8\deg$ is known as Cygnus X region.
           {\em Right:} Finding chart for OB associations (circles) and young 
           open clusters (asterisks) in the Cygnus region. For Cyg OB1 to 
           Cyg OB9 only the OB numbers are quoted.}
  \label{fig:maps}
\end{figure*}

\subsection{The Cygnus superbubble}
\label{sec:superbubble}

The discovery of an extended X-ray ring-like source surrounding the 
Cygnus X region by \cite{cash80} has been interpreted as a superbubble 
blown by the massive star winds and supernova explosions in the Cyg OB2 
association.
Anomalous stellar proper motions have been reported for stars in 
associations located near the edge of the bubble 
(Comer\'on et al. 1993\nocite{comeron93}), 
and have been interpreted as the sign for triggered star formation following 
gravitational instability of the expanding shell 
(Comer\'on \& Torra 1994\nocite{comeron94}).
However, the energetic requirements needed to explain the expansion 
motion are difficult to reconcile with the mechanical power provided 
by Cyg OB2 (Comer\'on et al. 1998\nocite{comeron98}).
In addition, the superbubble is equally well, if not even better, 
explained as a superposition of multiple objects at different 
distances that aline along the local spiral arm that is seen 
tangentially in this direction
(Bochkarev \& Sitnik 1985\nocite{bochkarev85};
Uyaniker et al. 2001\nocite{uyaniker01}).

Also our cluster analysis seems difficult to reconcile with the scenario of an 
expanding bubble that triggered formation of the associations 
surrounding Cyg OB2.
Firstly, the central association Cyg OB2 is one of the youngest in the 
area ($1-4$ Myr) while the surrounding associations Cyg OB1, OB3, 
OB7, and OB9 are slightly older ($2-6$ Myr).
In the scenario where Cyg OB2 triggered the formation of the other 
associations, however, the situation should be vice versa.
Secondly, only Cyg OB1 and OB3 are situated at distances that are 
compatible with that of Cyg OB2 while Cyg OB7 and OB9 are likely 
foreground objects.
In addition, Cyg OB2 is not situated at the centre of the X-ray superbubble 
and the other OB associations correlate only partly with the structure
(Uyaniker et al. 2001\nocite{uyaniker01}), making a physical relation 
of these objects highly questionable.
Thirdly, the expansion age determined by \cite{comeron98} from 
stellar proper motions fits surprisingly well our age estimate for Cyg 
OB2, supporting the idea that the expanding stars are dynamically 
ejected runaway stars from Cyg OB2 (Comer\'on et al. 1998\nocite{comeron98}).
Comer\'on et al.'s list of candidate members of the Cygnus expanding structure
contains four O-type stars, which compared to the total of 120 O-type 
stars in Cyg OB2 (Kn\"odlseder 2000\nocite{knoedl00}) 
would result in a runaway star fraction of $3\%$ for this association.
Compared to other young massive star associations, such as the 
Trapezium cluster in Orion or the $\lambda$ Orionis star-forming region
(Hoogerwerf et al. 2001\nocite{hoogerwerf01}), such a proportion seems 
typical, and we believe that at least part, if not all, of the anomalous 
stellar proper motions in Cygnus are indeed explained by runaway stars.

\subsection{First estimates}
\label{sec:first}
 
Using simple approximations we can already make a first order estimate 
of the ionising flux that we expect from the stellar populations in 
the Cygnus region.
The CoStar models of \cite{schaerer97} predict a Lyman continuum luminosity of 
$\log Q_{0} = 49.05$ \flyc\ for a star of spectral type O7V, and assuming 
that all 204 O stars that we find in the associations of the Cygnus region are 
indeed O7V stars, we derive a total luminosity of $\log Q = 51.36$ \flyc.
Assuming further that all stars lie at a typical distance of $1.8$ kpc, and 
using the relation
\begin{equation}
 \flxff\ (\mbox{Jy}) = 8.06 \times 10^{-48}\
 \frac{Q}{s^{2}}
 \left( \frac{T_{e}}{\mbox{8000 K}} \right)^{-0.45}
 \label{eq:ffconv}
\end{equation}
($s$ is the distance in units of kpc, $Q$ is the Lyc luminosity 
given in \flyc, and $T_{e}$ is the electron temperature which we 
assume to 8000 K; c.f. Afflerbach et al. 1997) we predict a 53 GHz 
free-free emission flux of $\sim5700$ Jy.
We will see in the next section that this flux level is of the same 
order as the observed value.

\section{Modelling the Cygnus region}
\label{sec:results}

\subsection{Gamma-ray and microwave observations}
\label{sec:observations}

Before we present the results of our evolutionary synthesis modelling 
we want to discuss the gamma-ray and microwave observations that 
triggered our investigation.
Maps of 1.809 \MeV\ gamma-ray line emission and 53 GHz free-free emission 
of the Cygnus regions are shown in superposition in Fig.~\ref{fig:maps}.
The spatial correlation between both emission features is striking.
However, we want to caution the reader of overinterpretations of the 
apparent structures in the map.
The gamma-ray feature has a total significance of only $\sim 8\sigma$, 
and the detailed morphology of the image is heavily biased by instrumental 
noise (Kn\"odlseder et al.~1999b\nocite{knoedl99b}).
Formally, the COMPTEL instrument had an angular resolution of 
$\sim4\deg$ (FWHM), yet the small amplitude of the signal probably 
does not allow inferences at angular scales below $\sim10\deg$.
This is comparable to the angular resolution of $7\deg$ (FWHM) for 
the {\em COBE} DMR experiment which was used to derive the free-free 
emission map.

Observations of galactic radio-continuum emission from Cygnus at smaller 
angular scales indicate that most of the emission originates from a region 
between $l\sim74\deg$ and $84\deg$ with emission peaks around $l\sim80\deg$
(e.g.~Wendker 1970\nocite{wendker70}).
This region coincides with the OB associations Cyg OB1, Cyg OB2, Cyg OB8, 
and Cyg OB9 and a large number of the youngest open clusters in our 
database.
However, the small-scale radio maps that are available were generally 
obtained at lower frequencies where synchrotron emission contributes 
significantly to the signal.
Although the Cygnus emission is primarily optically thin thermal 
radiation, a superimposed smooth non-thermal component makes the 
determination of absolute free-free emission levels difficult
(Wendker et al.~1991\nocite{wendker91}).
Despite the poor angular resolution, we therefore prefer using the 53 GHz 
free-free emission map of \cite{bennett92} for the estimation of the ionising 
flux in Cygnus.
At this frequency the contribution from synchrotron and thermal dust emission 
is small within the galactic plane, and can easily be removed by 
modelling their spatial distribution at adjacent frequencies.
The image shown in Fig.~\ref{fig:maps} results from such a separation 
performed on DMR data (Bennett et al.~1992\nocite{bennett92}).

Smaller angular scales are probably even not desirable for a global 
flux comparison, although they bear valuable information about the 
interplay between the massive stars with the interstellar medium.
Firstly, at angular scales that are smaller than the associations, 
the flux distribution may be influenced by individual massive stars, 
which we do not attempt to model in our study.
Secondly, the correlation between free-free emission and the massive 
star distribution may anyway break-down at scales comparable to the 
association dimensions due to the lack of gas in their ploughed interiors.
For example, \cite{huchtmeier77} did not find any \HII\ region inside 
Cyg OB2 despite the large number of ionising stars that is present.
They argue that the association is devoid of hydrogen and the ionising 
flux is converted into free-free emission by the surrounding gas, 
leading to a diffuse extended emission component.
Thirdly, the long lifetime of \al\ ($\tau \sim 1$ Myr) allows the 
isotope to travel substantial distances before the emission of the 
gamma-ray decay photons, and the detailed 1.809 \MeV\ emission 
structure should depend considerably on the interaction and 
deceleration of the ejecta in the surrounding interstellar medium.
Morphology studies that address the question of ejecta propagation in 
the Cygnus region will become possible with the upcoming 
{\em INTEGRAL} observatory 
(e.g.~Kn\"odlseder \& Vedrenne 2001\nocite{knoedl01}), but the 
existing COMPTEL data limits our analysis to only a global investigation.

We determined the 1.809 \MeV\ and 53 GHz free-free fluxes from the 
Cygnus region by integrating the respective skymaps over the area of 
interest.
A diffuse galactic ridge emission that underlies the features in the 
Cygnus region has been subtracted from the flux estimates in order to 
extract the emission that is correlated to the Cygnus associations. 
The resulting flux estimates are only weakly sensitive on the precise 
location of the integration boundaries.
To safely cover the investigated region we selected a longitude interval 
from $l=65\deg$ to $110\deg$ for the flux determination and 
determined the galactic ridge background from two adjacent intervals 
at $l=110\deg-130\deg$ and $l=60\deg-65\deg$.
We selected a rather large latitude window of $|b|<30\deg$ to assure 
that we recover also the tails of the emission (in particular the 
COMPTEL skymap shows extended high latitude wings due to the low 
significance of the reconstructed emission).
The resulting 1.809 \MeV\ gamma-ray line flux amounts to
$\flxal = (5.8\pm1.5) \times 10^{-5}$ \funit\ while the 53 GHz 
free-free emission flux equals $\flxff = 4200 \pm 700$ Jy.
The quoted errors reflect both the statistical and the systematical 
uncertainties in the flux determination, in particular those 
introduced by the selection of the integration boundaries.

The flux measurements can be converted into the equivalent O7V star 
\al\ yield using
\begin{equation}
 \yal\ (\Msol) = 7.91 \times 10^{3}\
 \frac{\flxal \, (\funit)}{\flxff \, (\mbox{Jy})} .
 \label{eq:y26conv}
\end{equation}
We obtain $\yal = (1.1 \pm 0.3) \times 10^{-4}$ \Msol\ for the Cygnus 
region, a value that is compatible with the galactic average value of 
$(1.0 \pm 0.3) \times 10^{-4}$ \Msol.
The flux measurements are summarised together with the equivalent \al\ 
yields in Table \ref{tab:flux}.

\begin{table}[t!]
  \footnotesize
  \caption{\label{tab:flux}
  Flux measurements and derived equivalent O7V star \al\ yield of 
  the Cygnus region. For comparison, the galactic average \yal\ value is 
  also quoted.
  }
  \begin{flushleft}
    \begin{tabular}{ll}
    \hline
    \noalign{\smallskip}
    Quantity & Measured value \\
    \noalign{\smallskip}
    \hline
    \noalign{\smallskip}
    \flxal         & $(5.8\pm1.5) \times 10^{-5}$ \funit \\
    \flxfe         & - \\
    \flxff         & $4200 \pm 700$ Jy \\
    \yal\          & $(1.1 \pm 0.3) \times 10^{-4}$ \Msol\ \\
    \noalign{\smallskip}
    \yal\ (Galaxy) & $(1.0 \pm 0.3) \times 10^{-4}$ \Msol\ \\
    \noalign{\smallskip}
    \hline
    \end{tabular}
  \end{flushleft}
\end{table}

\subsection{Evolutionary synthesis modelling}
\label{sec:synthesis}

\subsubsection{Prior probabilities and model uncertainties}
\label{sec:prior}

\begin{table*}[th]
  \footnotesize
  \caption{\label{tab:obresults}
  Results of the evolutionary synthesis calculations.
  The quoted values are the medians of the posterior probability 
  density functions, and the (asymmetric) errors specify the central
  $63.8\%$ confidence intervals.
  }
  \begin{flushleft}
    \renewcommand{\arraystretch}{1.3}
    \begin{tabular}{lccccccccccccc}
    \hline
    \noalign{\smallskip}
    Name & 
    \multicolumn{1}{c}{$\log Y_{26}$} &
    \multicolumn{1}{c}{$\log Y_{60}$} &
    \multicolumn{1}{c}{$\log Q$} &
    \multicolumn{1}{c}{$\log \flxal$} &
    \multicolumn{1}{c}{$\log \flxfe$} &
    \multicolumn{1}{c}{$\log \flxff$} &
    \multicolumn{1}{c}{$N_{O}$} &
    \multicolumn{1}{c}{$N_{WR}$} \\
    &
    \multicolumn{1}{c}{(\Msol)} &
    \multicolumn{1}{c}{(\Msol)} &
    \multicolumn{1}{c}{(\flyc)} &
    \multicolumn{1}{c}{(\funit)} &
    \multicolumn{1}{c}{(\funit)} &
    \multicolumn{1}{c}{(Jy)} &
    &
    \\
    \noalign{\smallskip}
    \hline
    \noalign{\smallskip}
    Cep OB1a    & $-3.2^{+0.2}_{-0.4}$             
                & $-\infty$                        
                & $49.8^{+0.5}_{-0.4}$             
                & $-6.3^{+0.3}_{-0.4}$             
                & $-\infty$                        
                & $1.7^{+0.5}_{-0.4}$              
                & $12.7^{+6.5}_{-12.7}$            
                & $1.8^{+2.5}_{-1.7}$ \\           
    Cep OB1b    & $-4.3^{+0.3}_{-0.4}$             
                & $-3.9^{+0.2}_{-0.3}$             
                & $46.8^{+0.5}_{-0.5}$             
                & $-7.3^{+0.4}_{-0.4}$             
                & $-7.6^{+0.3}_{-0.3}$             
                & $-1.4^{+0.5}_{-0.5}$             
                & $0$                              
                & $0$ \\                           
    Cep OB2     & $-4.8^{+0.5}_{-0.9}$             
                & $-4.4^{+0.3}_{-0.5}$             
                & $46.4^{+0.4}_{-0.4}$             
                & $-6.3^{+0.5}_{-0.9}$             
                & $-6.6^{+0.3}_{-0.5}$             
                & $-0.3^{+0.4}_{-0.4}$             
                & $0$                              
                & $0$ \\                           
    Cyg OB1     & $-3.4^{+0.3}_{-0.3}$             
                & $-3.8^{+0.5}_{-\infty}$          
                & $49.6^{+0.5}_{-0.7}$             
                & $-5.9^{+0.3}_{-0.4}$             
                & $-7.0^{+0.5}_{-\infty}$          
                & $2.0^{+0.6}_{-0.7}$              
                & $5.6^{+8.7}_{-5.6}$              
                & $1.0^{+2.3}_{-1.0}$ \\           
    Cyg OB2     & $-2.6^{+0.2}_{-1.0}$             
                & $-\infty$                        
                & $51.0^{+0.2}_{-0.6}$             
                & $-5.0^{+0.4}_{-1.0}$             
                & $-\infty$                        
                & $3.5^{+0.3}_{-0.5}$              
                & $97.1^{+18.7}_{-46.6}$           
                & $8.9^{+8.9}_{-8.9}$ \\           
    Cyg OB3     & $-3.3^{+0.2}_{-0.4}$             
                & $-\infty$                        
                & $49.9^{+0.5}_{-0.3}$             
                & $-5.8^{+0.4}_{-0.5}$             
                & $-\infty$                        
                & $2.2^{+0.5}_{-0.5}$              
                & $12.4^{+7.9}_{-12.4}$            
                & $1.9^{+2.5}_{-1.8}$ \\           
    Cyg OB7     & $-7.3^{+3.3}_{-\infty}$          
                & $-\infty$                        
                & $48.7^{+0.6}_{-0.7}$             
                & $-9.0^{+3.3}_{-\infty}$          
                & $-\infty$                        
                & $1.8^{+0.6}_{-0.7}$              
                & $0.4^{+1.8}_{-0.4}$              
                & $0.0^{+0.3}_{}$ \\               
    Cyg OB8     & $-4.0^{+0.3}_{-0.6}$             
                & $-3.8^{+0.2}_{-\infty}$          
                & $47.7^{+1.8}_{-1.1}$             
                & $-6.6^{+0.4}_{-0.6}$             
                & $-7.2^{+0.4}_{-\infty}$          
                & $-0.1^{+1.8}_{-1.1}$             
                & $0.0^{+5.7}_{}$                  
                & $0.0^{+0.3}_{}$ \\               
    Cyg OB9     & $-3.7^{+0.4}_{-1.2}$             
                & $-\infty$                        
                & $49.4^{+0.5}_{-0.4}$             
                & $-5.8^{+0.4}_{-1.2}$             
                & $-\infty$                        
                & $2.1^{+0.5}_{-0.5}$              
                & $4.4^{+2.7}_{-4.4}$              
                & $0.3^{+1.4}_{-0.3}$ \\           
    Lac OB1     & $-4.5^{+0.3}_{-0.5}$             
                & $-4.2^{+0.3}_{-0.4}$             
                & $46.6^{+0.3}_{-0.2}$             
                & $-6.0^{+0.4}_{-0.6}$             
                & $-6.4^{+0.4}_{-0.4}$             
                & $-0.1^{+0.3}_{-0.3}$             
                & $0$                              
                & $0$ \\                           
    Vul OB1     & $-4.2^{+0.5}_{-\infty}$          
                & $-\infty$                        
                & $48.8^{+0.6}_{-0.7}$             
                & $-6.9^{+0.5}_{-\infty}$          
                & $-\infty$                        
                & $1.0^{+0.6}_{-0.7}$              
                & $0.3^{+2.6}_{-0.3}$              
                & $0.0^{+0.7}_{}$ \\               
    \noalign{\smallskip}
    \hline
    \noalign{\smallskip}
    Berkeley 86 & $-4.3^{+0.5}_{-\infty}$          
                & $-\infty$                        
                & $48.7^{+0.5}_{-0.7}$             
                & $-6.7^{+0.6}_{-\infty}$          
                & $-\infty$                        
                & $1.2^{+0.5}_{-0.7}$              
                & $0.1^{+1.9}_{-0.1}$              
                & $0.0^{+0.6}_{}$ \\               
    Berkeley 87 & $-3.9^{+0.4}_{-0.6}$             
                & $-4.2^{+0.5}_{-\infty}$          
                & $48.9^{+0.4}_{-0.7}$             
                & $-6.4^{+0.6}_{-0.8}$             
                & $-7.5^{+0.7}_{-\infty}$          
                & $1.3^{+0.7}_{-0.7}$              
                & $0.0^{+3.3}_{}$                  
                & $0.0^{+1.0}_{}$ \\               
    Berkeley 94 & $-\infty$                        
                & $-\infty$                        
                & $48.2^{+0.7}_{-2.0}$             
                & $-\infty$                        
                & $-\infty$                        
                & $-0.4^{+0.7}_{-1.9}$             
                & $0.0^{+0.8}_{}$                  
                & $0.0^{+0.2}_{}$ \\               
    Berkeley 96 & $-\infty$                        
                & $-\infty$                        
                & $47.2^{+0.9}_{-1.2}$             
                & $-\infty$                        
                & $-\infty$                        
                & $-1.3^{+0.9}_{-1.2}$             
                & $0$                              
                & $0$ \\                           
    Biukaran 2  & $-6.6^{+1.3}_{-2.5}$             
                & $-5.7^{+0.5}_{-1.0}$             
                & $44.8^{+0.2}_{-0.2}$             
                & $-9.2^{+1.3}_{-2.5}$             
                & $-9.0^{+0.5}_{-1.0}$             
                & $-2.9^{+0.3}_{-0.3}$             
                & $0$                              
                & $0$ \\                           
    IC 4996     & $-4.8^{+0.7}_{-\infty}$          
                & $-4.8^{+0.7}_{-\infty}$          
                & $47.6^{+0.8}_{-0.7}$             
                & $-7.2^{+0.7}_{-\infty}$          
                & $-7.9^{+0.7}_{-\infty}$          
                & $0.1^{+0.8}_{-0.8}$              
                & $0$                              
                & $0$ \\                           
    NGC 6823    & $-4.1^{+0.6}_{-\infty}$          
                & $-\infty$                        
                & $49.3^{+0.6}_{-0.6}$             
                & $-6.9^{+0.7}_{-\infty}$          
                & $-\infty$                        
                & $1.4^{+0.6}_{-0.7}$              
                & $2.9^{+3.2}_{-2.9}$              
                & $0.0^{+0.9}_{}$ \\               
    NGC 6871    & $-4.2^{+0.3}_{-\infty}$          
                & $-4.3^{+0.4}_{-\infty}$          
                & $48.3^{+0.6}_{-0.5}$             
                & $-6.9^{+0.4}_{-\infty}$          
                & $-7.6^{+0.4}_{-\infty}$          
                & $0.4^{+0.6}_{-0.6}$              
                & $0$                              
                & $0.0^{+0.3}_{}$ \\               
    NGC 6883    & $-4.9^{+0.4}_{-0.7}$             
                & $-4.3^{+0.3}_{-0.6}$             
                & $46.1^{+0.3}_{-0.6}$             
                & $-7.3^{+0.5}_{-0.8}$             
                & $-7.5^{+0.5}_{-0.6}$             
                & $-1.5^{+0.5}_{-0.6}$             
                & $0$                              
                & $0$ \\                           
    NGC 6910    & $-7.1^{+3.1}_{-\infty}$          
                & $-\infty$                        
                & $48.3^{+0.5}_{-0.9}$             
                & $-9.6^{+3.2}_{-\infty}$          
                & $-\infty$                        
                & $0.7^{+0.5}_{-0.9}$              
                & $0.0^{+0.7}_{}$                  
                & $0$ \\                           
    NGC 6913    & $-4.8^{+1.1}_{-\infty}$          
                & $-\infty$                        
                & $48.8^{+0.6}_{-0.7}$             
                & $-7.3^{+1.3}_{-\infty}$          
                & $-\infty$                        
                & $1.2^{+0.8}_{-0.8}$              
                & $0.5^{+1.9}_{-0.5}$              
                & $0.0^{+0.6}_{}$ \\               
    NGC 7067    & $-4.7^{+0.3}_{-0.5}$             
                & $-4.0^{+0.3}_{-0.4}$             
                & $46.3^{+0.2}_{-0.2}$             
                & $-8.2^{+0.3}_{-0.5}$             
                & $-8.3^{+0.3}_{-0.4}$             
                & $-2.4^{+0.2}_{-0.2}$             
                & $0$                              
                & $0$ \\                           
    NGC 7128    & $-5.5^{+0.6}_{-1.2}$             
                & $-5.2^{+0.3}_{-0.5}$             
                & $45.2^{+0.2}_{-0.2}$             
                & $-8.6^{+0.6}_{-1.2}$             
                & $-9.1^{+0.3}_{-0.5}$             
                & $-3.2^{+0.3}_{-0.3}$             
                & $0$                              
                & $0$ \\                           
    NGC 7160    & $-5.4^{+0.9}_{-\infty}$          
                & $-4.8^{+0.5}_{-\infty}$          
                & $46.2^{+0.6}_{-1.1}$             
                & $-7.3^{+0.9}_{-\infty}$          
                & $-7.4^{+0.5}_{-\infty}$          
                & $-0.9^{+0.6}_{-1.1}$             
                & $0$                              
                & $0$ \\                           
    NGC 7235    & $-3.8^{+0.3}_{-0.7}$             
                & $-4.2^{+0.4}_{-\infty}$          
                & $49.1^{+0.3}_{-0.4}$             
                & $-6.9^{+0.4}_{-0.7}$             
                & $-7.9^{+0.4}_{-\infty}$          
                & $0.9^{+0.4}_{-0.5}$              
                & $0.6^{+2.6}_{-0.6}$              
                & $0.1^{+1.0}_{-0.1}$ \\           
    NGC 7261    & $-5.1^{+0.5}_{-1.2}$             
                & $-4.6^{+0.4}_{-0.6}$             
                & $46.0^{+0.4}_{-0.4}$             
                & $-8.1^{+0.6}_{-1.2}$             
                & $-8.3^{+0.5}_{-0.7}$             
                & $-2.2^{+0.4}_{-0.4}$             
                & $0$                              
                & $0$ \\                           
    NGC 7380    & $-4.5^{+0.6}_{-\infty}$          
                & $-5.7^{+1.5}_{-\infty}$          
                & $48.3^{+0.6}_{-0.8}$             
                & $-7.5^{+0.6}_{-\infty}$          
                & $-9.4^{+1.5}_{-\infty}$          
                & $0.1^{+0.6}_{-0.8}$              
                & $0.0^{+0.7}_{}$                  
                & $0.0^{+0.2}_{}$ \\               
    Roslund 4   & $-4.9^{+0.6}_{-1.6}$             
                & $-4.5^{+0.4}_{-1.0}$             
                & $46.5^{+1.0}_{-0.6}$             
                & $-7.7^{+0.7}_{-1.6}$             
                & $-8.1^{+0.5}_{-1.0}$             
                & $-1.4^{+1.0}_{-0.7}$             
                & $0$                              
                & $0$ \\                           
    Roslund 5   & $-8.5^{+2.7}_{-\infty}$          
                & $-6.4^{+1.0}_{-2.7}$             
                & $44.5^{+0.5}_{-1.0}$             
                & $-9.8^{+2.7}_{-\infty}$          
                & $-8.4^{+1.0}_{-2.8}$             
                & $-2.0^{+0.5}_{-1.0}$             
                & $0$                              
                & $0$ \\                           
    Ruprecht 175 & $-\infty$                       
                & $-7.6^{+1.9}_{-\infty}$          
                & $43.9^{+0.8}_{-1.9}$             
                & $-\infty$                        
                & $-10.8^{+1.9}_{-\infty}$         
                & $-\infty$                        
                & $0$                              
                & $0$ \\                           
    Trumpler 37 & $-3.9^{+0.4}_{-0.6}$             
                & $-4.2^{+0.5}_{-\infty}$          
                & $49.0^{+0.4}_{-0.7}$             
                & $-5.8^{+0.4}_{-0.6}$             
                & $-6.9^{+0.5}_{-\infty}$          
                & $1.8^{+0.5}_{-0.7}$              
                & $0.0^{+3.5}_{}$                  
                & $0.0^{+1.2}_{}$ \\               
    \noalign{\smallskip}
    \hline
    \noalign{\smallskip}
    total       & - & - & - 
                & $-4.6^{+0.2}_{-0.2}$
                & $-5.7^{+0.3}_{-0.1}$ 
                & $3.7^{+0.2}_{-0.4}$
                & $145.1^{+26.5}_{-46.0}$ 
                & $22.6^{+10.9}_{-9.7}$ \\
    \noalign{\smallskip}
    \hline
    \end{tabular}
  \end{flushleft}
\end{table*}

We used the method described in \cite{cervino00} and Sect.~\ref{sec:model} 
to estimate the nucleosynthesis yields and Lyc luminosity for each 
association.
The results are compiled in Tab.~\ref{tab:obresults}.
For each association 100 statistically independent evolutionary 
synthesis models have been calculated to approximate the PDFs for all 
quantities of interest.
Based on our age determinations, we marginalised the age uncertainty 
using Eq.~\ref{eq:age} with bounded Gaussian-shaped prior probability 
distributions.
The mean and standard deviation of the Gaussians were defined by the 
mean and the half-width of the age boundaries quoted in column 2 of 
Tab.~\ref{tab:obdata}; the prior density was set to zero for ages outside 
the age boundaries.
The \al\ and \fe\ yields as well as the Lyc luminosities that result 
from this procedure are given in columns $2-4$ of Tab.~\ref{tab:obresults}.

Alternatively, we also tried box-shaped prior distributions that were 
bounded by our age estimates, but the results were only slightly 
different.
In fact, Eq.~\ref{eq:age} performs a weighted average of the PDFs 
where the age prior $p(t)$ defines the weights that are attributed to 
each age.
From our association analysis we derive a possible range of ages, yet 
the data are not sufficiently accurate to derive details about the 
age distribution.
Therefore, any prior distribution that reflects reasonably well the 
fact that the association age is comprised between two limits should 
provide a valid prior for our analysis.
The only difference between the Gaussian and the box-shaped prior is 
that the first slightly favours models with an age close to the mean 
age between the boundaries, while the second gives equal weight to all 
ages that are compatible with the assigned limits.
Since the observed age spread is at least partly due to measurement 
uncertainties, and since we defined our age boundaries as two extreme
limits that enclose the age distribution of the most luminous stars, 
we feel that a bounded Gaussian-shaped prior reflects most properly 
the information that we have at hand about the association ages.

We also want to emphasise that our age marginalisation is equivalent 
to the assumption of a continuous star formation over an interval 
defined by the age boundaries.
In this interpretation, the form of the age prior specifies the time 
evolution of the star formation rate.
Since this rate is generally unknown, a rather flat prior 
distribution, like those we discussed above, is certainly appropriate.
The interpretation of the apparent age distribution in the H-R diagram 
is usually a difficult enterprise due to the blending between 
measurement uncertainties, spectral misclassification, neglection of 
binarity and rotation, non-member pollution, and a possible non-coeval 
star formation.
We cannot distinguish between all these uncertainties either, yet our 
formalism includes all of them into the model results.

The nucleosynthesis yields and ionising luminosity were converted into 
gamma-ray line and microwave free-free fluxes by marginalising the 
distance uncertainty for each association (Eq.~\ref{eq:distance}).
For the 1.809 \MeV\ gamma-ray line flux due to the radioactive decay 
of \al\ we used the relation
\begin{equation}
 \flxal\ (\funit) = 1.18 \times 10^{-2}\ \frac{Y_{26}}{s^2}
\end{equation}
while for the 1.137 \MeV\ gamma-ray line flux from the decay of \fe\ we 
employed
\begin{equation}
 \flxfe\ (\funit) = 2.41 \times 10^{-3}\ \frac{Y_{60}}{s^2}
\end{equation}
($Y_{26}$ and $Y_{60}$ are the \al\ and \fe\ yields in units of \Msol, 
respectively, $s$ is the distance in units of kpc).
Note that the \fe\ decay leads in fact to two gamma-ray lines at 
1.137 and 1.332 \MeV, but since their branching ratio and thus their 
intensity is basically identical we only quote the flux for the 1.137 
\MeV\ line.
The conversion from ionising luminosity to 53 GHz free-free flux is 
given by Eq.~\ref{eq:ffconv}.
As distance priors $p(s)$ we used Gaussians with means and standard 
deviations that were defined by the distance moduli means and errors 
given in column 3 of Tab.~\ref{tab:distance}.
Again, since the distance modulus uncertainty arises primarily from 
measurement uncertainties, we believe that a Gaussian-shaped prior is 
a reasonable representation of our distance knowledge.
The resulting 1.809 \MeV, 1.137 \MeV, and 53 GHz fluxes are quoted in 
columns $5-7$ of Tab.~\ref{tab:obresults}.

Our evolutionary synthesis model also predicts the spectral type 
distribution, and we quote in columns $8-9$ of Tab.~\ref{tab:obresults} 
the predicted number of O-type and Wolf-Rayet stars.
These predictions can be directly compared to the respective columns 
of Tab.~\ref{tab:obdata}, and we globally find a good agreement 
although the median values often fall below the observed values.
The exception to the rule is Cyg OB2 for which the models predict 
between 51 and 116 O stars while the observations suggest 120.
However, as discussed in Sect.~\ref{sec:masses}, the 120 O stars 
quoted in Tab.~\ref{tab:obdata} are rather an upper limit since the 
{\em 2MASS} analysis did not account for possible evolutionary 
effects.
Yet, Cyg OB2 houses evolved stars and even shows indications for 
non-coeval and ongoing star formation
(e.g. Torres-Dodgen et al. 1991\nocite{torres91}; 
Massey \& Thompson 1991\nocite{massey91};
Parthasarathy et al. 1992\nocite{parthasarathy92};
Pigulski \& Kolaczkowski 1998\nocite{pigulski98};
Herrero et al. 1999\nocite{herrero99}).

\begin{figure}[tb]
  \includegraphics[width=8.8cm]{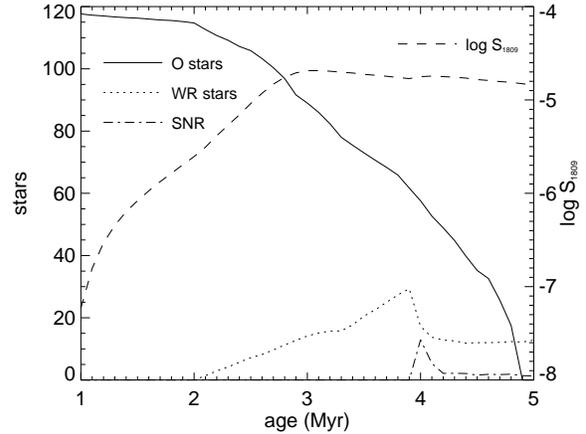}
  \caption{Evolution of the number of O stars, Wolf-Rayet stars, 
           supernova remnants, and 1.809 \MeV\ gamma-ray line flux as 
           function of age for Cyg OB2. We assumed a lifetime of $10^5$ yr 
           for each SNR.}
  \label{fig:CygOB2age}
\end{figure}

The effect of the age uncertainty on the stellar population in Cyg 
OB2 is illustrated in Fig.~\ref{fig:CygOB2age}.
For a coeval star formation event, the number of O stars drops rapidly 
from the initial value of 120 to zero during the first 5 Myr of the 
association evolution.
Clearly, there is a considerable number of O-type stars in Cyg OB2 
(e.g. Massey et al. 1991\nocite{massey91}), hence the associations 
should be younger than 5 Myr.
Wolf-Rayet stars appear around 2 Myr after the starburst while the first 
supernovae explode about 4 Myr after the birth of the association.
The fact that there are Wolf-Rayet stars associated with Cyg OB2 
suggests an age $\ga2$ Myr, the presence of 2 WC-type stars even 
indicates an age above 3 Myr.
On the other hand no supernova remnant has been detected within the 
boundaries of Cyg OB2 (Wendker et al. 1991\nocite{wendker91}), hence 
the association should be younger than $\sim4$ Myr.
The 1.809 \MeV\ gamma-ray line flux raises from zero to a maximum level 
of $2\times10^{-5}$ \funit\ within $\sim3$ Myr (for an assumed distance 
modulus of $DM=11.0$ mag), and Cyg OB2 could be situated just at the maximum 
of the 1.809 \MeV\ gamma-ray light curve, which is about a factor of two 
above the median value quoted in Tab.~\ref{tab:obresults}.
Nevertheless, the present reasoning only holds for a strictly coeval 
star formation event, and the evidence for ongoing star formation in 
Cyg OB2 holds against narrowing down the age boundaries for the 
association.
Clearly, more observational data is required to improve our knowledge 
about the age distribution in Cyg OB2 before we safely can improve 
upon the nucleosynthesis yield estimation for this association.

This example illustrates that the apparent age spread in the HRDs presents 
an important source of uncertainty for the final flux estimates, at least 
for the youngest associations.
In general, however, the dominating source of uncertainty depends on 
the association richness and its age, but also on the nature of the 
quantity of interest.
For Cyg OB2, for example, the 53 GHz flux prediction is much less 
affected by the age uncertainty than the 1.809 \MeV\ flux estimate.
While the 1.809 \MeV\ flux varies by more than two magnitudes between 
$1-4$ Myr, the number of O stars changes only by a factor of two, 
implying a 53 GHz flux variation by a factor of less than six.
For slightly older associations, such as Trumpler 37, the situation 
is reversed: the 1.809 \MeV\ flux varies only slowly 
while the ionising flux now drops quickly due to the disappearing of 
the most massive stars in supernovae events.

If the association is sufficiently rich so that sampling effects are 
negligible, the distance uncertainty turns out to be the dominating 
factor of the flux dispersion.
Typically, the relative distance uncertainty for an association 
amounts to $20\%$ which directly translates into a flux uncertainty 
of $40\%$.
Sampling effects are important when only few stars contribute to the 
association luminosities, i.e. either for very young or for very poor 
populations.
Cyg OB7 fulfils both conditions.
For this association, a considerable fraction of the evolutionary synthesis 
samples did not lead to any \al\ production at all.
In those cases, the initial mass function has not been populated by 
sufficiently massive stars to lead to a noticeable \al\ production 
during the conceivable age range of Cyg OB7 ($2-5$ Myr).
We indicate such a possibility by quoting infinite negative errors in 
Tab.~\ref{tab:obresults}.

\subsubsection{Correlated emission}

\begin{figure}[tb]
  \includegraphics[width=8.8cm]{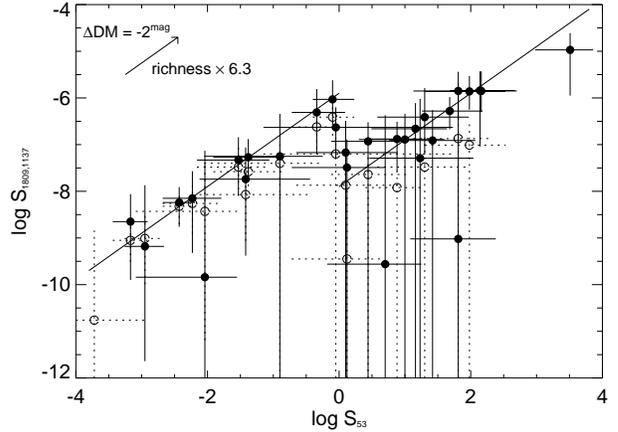}
  \caption{Predicted 1.809 \MeV\ (filled dots and solid error bars) and 
           1.137 \MeV\ gamma-ray line fluxes (open dots and dashed error 
           bars) as function of predicted 53 GHz free-free flux.
           The arrow indicates the displacement due to a decrease of $2$ mag 
           in the distance modulus or an increase of the richness by a 
           factor of $6.3$. The two solid lines indicate equivalent \al\ 
           O7V star yields of $\log \yal = -2$ (left) and $-4$ (right).}
  \label{fig:correlation}
\end{figure}

To investigate the correlation between gamma-ray line flux and 
free-free emission in the Cygnus region we plotted the predicted 
1.809 and 1.137 \MeV\ fluxes versus the 53 GHz flux 
(Fig.~\ref{fig:correlation}).
All three quantities should scale with distance and association richness, 
and indeed we find a general trend of increasing gamma-ray intensity 
with increasing free-free flux.
A few associations with low gamma-ray fluxes 
($\log S_{1809,1137} \approx -9$) deviate 
from this tendency, yet they represent those poorly populated clusters 
that show a tremendous flux dispersion due to poor sampling of the 
high-mass IMF (see above).
To illustrate the distance and richness impact, we superimposed an 
arrow in Fig.~\ref{fig:correlation} that indicates the displacement 
due to a decrease of $2$ mag in the distance modulus or an equivalent
increase of the association richness by a factor of $6.3$.

For the 1.809 \MeV\ flux, the associations are situated along this line 
of displacement, yet there occurs a clear discontinuity around 
$\log \flxff \approx 0$.
All associations with $\log \flxff < 0$ are older than $10$ Myr, an 
evolutionary phase during which \al\ is exclusively ejected by supernovae 
events (the only exception to this rule are Berkeley 94 and 96, yet they are 
so sparsely populated that no \al\ production is expected).
In contrast, all associations with $\log \flxff > 0$ are younger 
than 10 Myr, hence they are dominated by \al\ ejection through stellar winds
(Cervi\~no et al. 2000\nocite{cervino00}).
The discontinuity in Fig.~\ref{fig:correlation} represents a change in 
the equivalent O7V star \al\ yield, and we find typical values of 
$\log \yal \approx -2$ and $-4$ for $\log \flxff < 0$ and $>0$, 
respectively.

The discontinuity in the correlation coefficient may eventually lead to 
a break-down in the tight correlation between 1.809 \MeV\ and 53 GHz 
emission, which should occur however at rather low 1.809 \MeV\ flux 
levels of about $10^{-6}$ \funit.
At this level, it is expected to find 1.809 \MeV\ emission features 
that are uncorrelated to microwave free-free emission (or for which 
the free-free emission should not exceed an intensity of 1 Jy).
Given the sensitivity limit of the spectrometer SPI on 
{\em INTEGRAL}, it is unlikely that this break-down will be 
observable in the near future.

For the 1.137 \MeV\ flux we find only a weak correlation with free-free 
emission, which is mainly limited to $\log \flxff < 0$, i.e. older 
associations that are dominated by supernova explosions.
For those, a typical equivalent O7V star \fe\ yield of
$\log \yfe \approx -1.5$ is found, yet the low level of the associated 
free-free flux ($\flxff < 1$ Jy) will inhibit the observation of this 
correlation.
For $\log \flxff > 0$, 1.137 \MeV\ fluxes are highly uncertain due to 
the small number of supernova events that contribute to \fe\ production
(recall that no large amounts of wind-ejected \fe\ is expected from 
massive stars; see Cervi\~no et al. 2000\nocite{cervino00}).
The youngest associations, such as Cyg OB2 or Cyg OB3, show no \fe\ 
production at all, despite the huge amounts of ionising photons that is 
produced by their large number of O-type stars.

By far the most luminous association in both 1.809 \MeV\ and 53 GHz 
free-free emission is Cyg OB2, which corresponds to the isolated dot 
in the upper-right corner of Fig.~\ref{fig:correlation}.
The median equivalent yield of Cyg OB2 amounts to 
$\yal = 2.2 \times 10^{-5}$ \Msol, somewhat lower than the observed value 
of $1.1 \times 10^{-4}$ \Msol\ in the Cygnus region.
In other words, Cyg OB2 seems to underproduce \al\ with respect to 
ionising photons, which is mainly related to its extremely young age. 
Most of the massive stars are still on the main sequence where they 
produce large amounts of ionising photons, but do not yet eject 
copious amounts of \al\ in the interstellar medium.

For the entire Cygnus region, we find an equivalent O7V star 
\al\ yield of $\yal = 4.7 \times 10^{-5}$ \Msol, about a factor of two 
below the observed value.
Apparently, our model underpredicts \al\ production with respect to 
the ionising luminosity by about a factor of 2, and we will suggest later 
that this underprediction is most likely related to shortcomings of 
current nucleosynthesis models for massive stars.
It is interesting to note that our predicted \yal\ value is very 
close to the steady-state value of $4.9 \times 10^{-5}$ \Msol, which 
results for a stellar population with a constant star formation rate
(Cervi\~no et al. 2000\nocite{cervino00}).
Since the Cygnus region is largely dominated by a single association, 
Cyg OB2, we believe that this finding is merely a coincidence.
Indeed, our predicted equivalent O7V star \fe\ yield of 
$\yfe = 2.0 \times 10^{-5}$ \Msol\ for the entire Cygnus region falls 
more than a factor of two below the steady-state value of
$5.6 \times  10^{-5}$ \Msol\ (Cervi\~no et al. 2000\nocite{cervino00}), 
emphasising the relative rareness of supernova events in the Cygnus region 
with respect to a steady-state population.

\subsubsection{Spatial distribution}

\begin{figure*}[tb]
  \includegraphics[width=8.9cm]{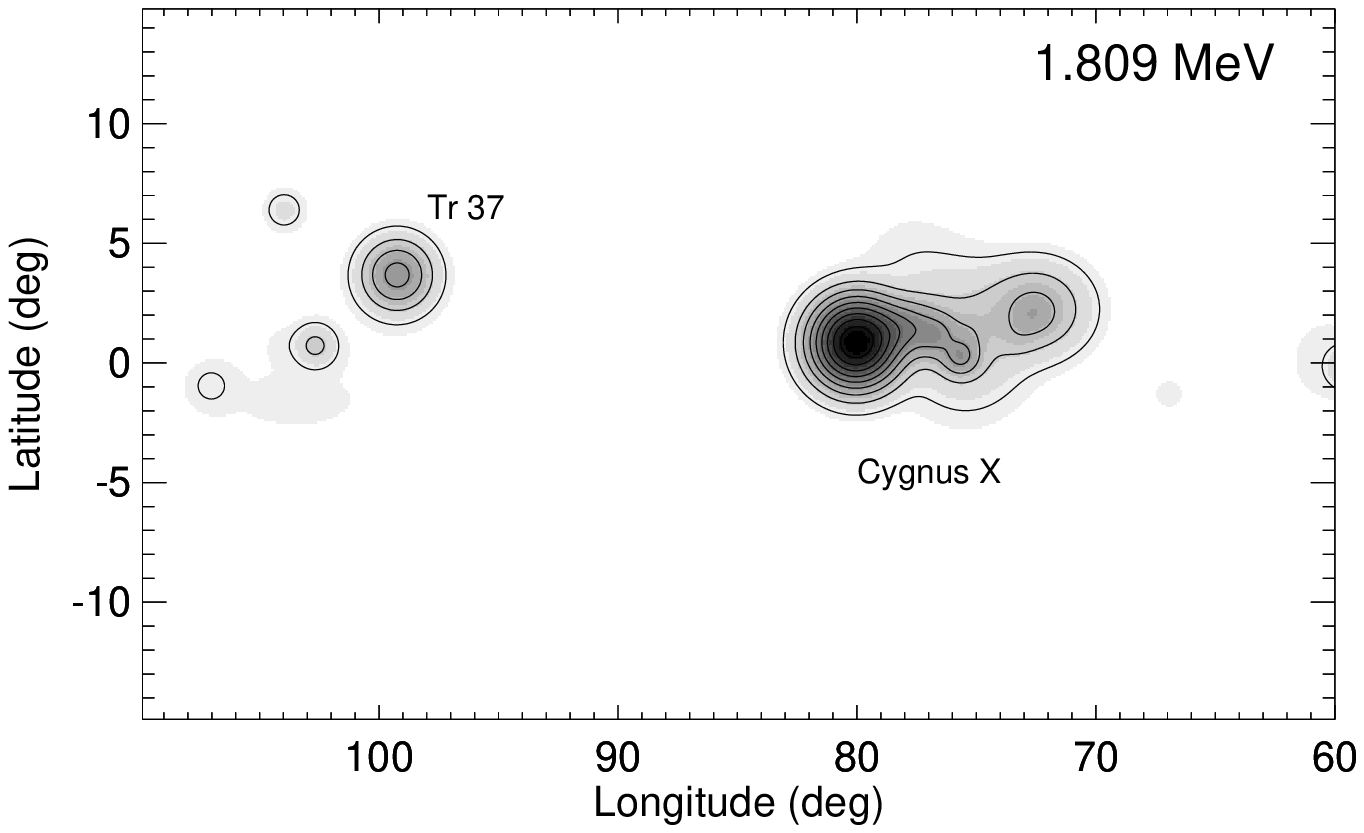} \hfill
  \includegraphics[width=8.9cm]{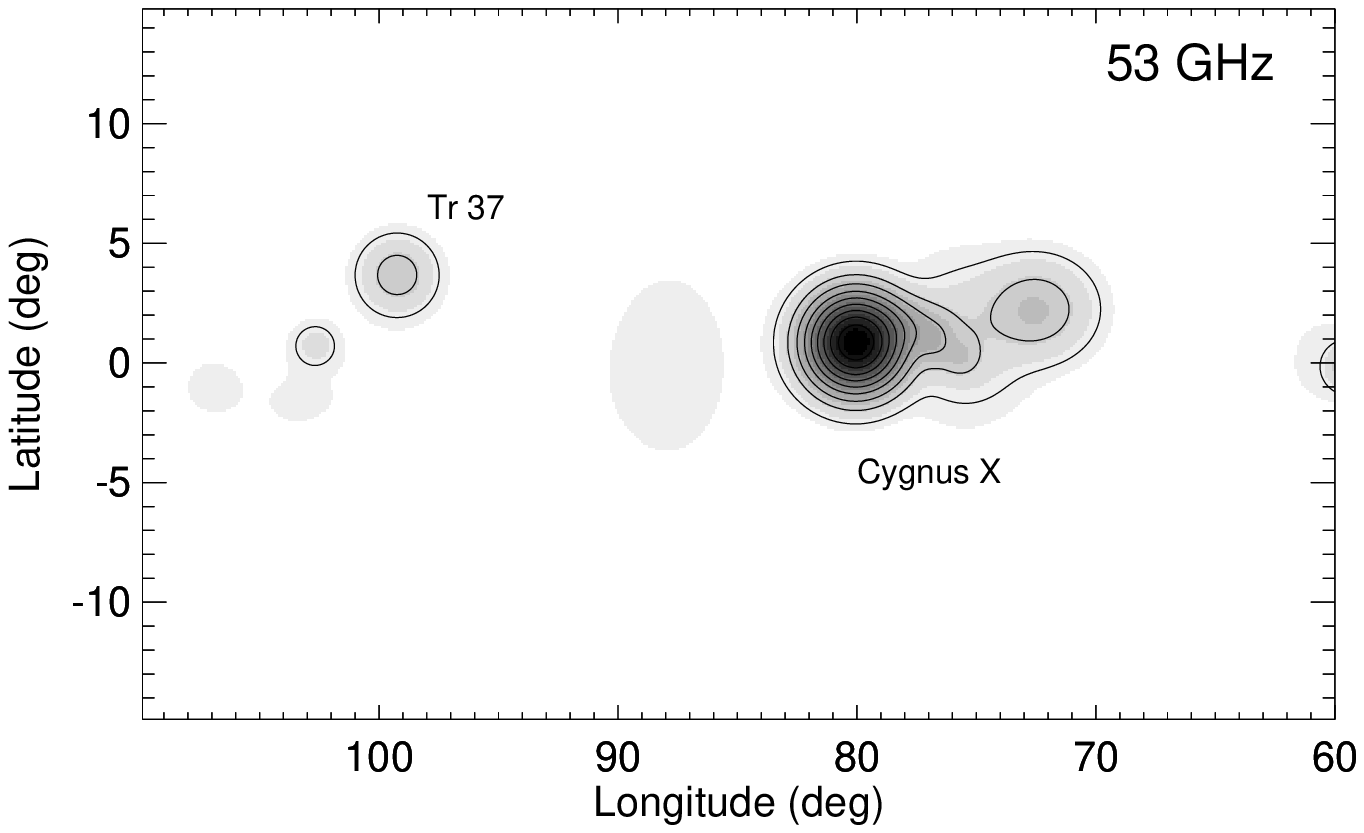}
  \caption{Predicted median 1.809 \MeV\ gamma-ray line intensity distribution 
           (left) and 53 GHz microwave free-free intensity (right).}
  \label{fig:modelmaps}
\end{figure*}

To illustrate the spatial distribution of the 1.809 \MeV\ and 53 GHz 
emission predicted by our model, we show in Fig.\ref{fig:modelmaps} intensity 
maps for both observables that are based on the median flux values quoted in 
Tab.~\ref{tab:obresults}.
Although we have no information about the spatial distribution of the 
respective emission within and around the associations, we 
distributed the flux within the association boundaries using 
Gaussian-shaped density profiles to account for the spatial extent of 
the stellar populations.
We recall that the true distribution could be more widespread due to 
the possible lack of ionised gas in the ploughed cluster interiors and 
the propagation of the radioactive ejecta before decay 
(cf. Sect.~\ref{sec:observations}).
In addition, when comparing these distributions to the observed maps 
in Fig.~\ref{fig:maps}, the angular resolution of the telescopes of
$4\deg$ (COMPTEL) and $7\deg$ (DMR) has to be considered.

Taking these factors into account, the morphological resemblance of 
our model maps to the observations is quite reasonable.
We predict free-free and 1.809 \MeV\ emission maxima at 
$(l,b)\sim(80\deg,1\deg)$ which spatially coincides with the observed 
maximum of the 53 GHz map, and is also close the maxima in the 1.809 
\MeV\ COMPTEL map.
The observations show copious emission in the longitude range 
$l\sim70\deg - 90\deg$, and also our model predicts the bulk emission 
in this region.
Obviously, the most prolific sources of 1.809 \MeV\ and 53 GHz 
emission are (in the quoted order) the OB associations Cyg OB2, Cyg OB3, 
Cyg OB9, and Cyg OB1 that form the core of the Cygnus X complex (see 
Fig.~\ref{fig:maps}).
Cyg OB2 is by far the most important source in the area for which we 
predict median fluxes of $\flxal = 1\times10^{-5}$ \funit\ and 
$\flxff = 3200$ Jy.
Comparison of the median values with the observed fluxes from the 
entire region (cf. Sect.~\ref{sec:observations}) suggests that Cyg OB2 
alone could provide around $75\%$ of the ionisation in Cygnus while 
it may account for only $\sim20\%$ of the observed 1.809 \MeV\ photons.
Including both observational and modelling uncertainties weakens this 
discrepancy, yet the general trend remains:
within the uncertainties, Cyg OB2 could easily explain all ionisation 
in Cygnus while it can account at most for $60\%$ of the 
observed \al.

Using successive marginalisation (Eq.~\ref{eq:total}) we estimate the 
total flux from the associations and clusters in the Cygnus X area
($70\deg \le l \le 82\deg$; $-3\deg \le b \le 6\deg$) 
to 
$\flxal = (2.0_{-1.1}^{+1.6}) \times 10^{-5}$ \funit,
$\flxfe = (3.4_{-2.6}^{+6.1}) \times 10^{-8}$ \funit, 
and $\flxff = 4000_{-2300}^{+3700}$ Jy.
The ionising flux is in comfortable agreement with the integrated 53 
GHz emission from the Cygnus region, while the 1.809 \MeV\ flux 
falls significantly below the observed value.
The predicted 1.137 \MeV\ flux is way below the detection limit of any 
envisagable gamma-ray telescope, reflecting the extreme rareness of 
supernova events in this area. 
Stellar--wind ejection seems to dominate the enrichment of the 
interstellar medium by nucleosynthesis products in the Cygnus X 
region -- more than $85\%$ of the 1.809 \MeV\ flux in our model 
originates from \al\ that is ejected through this channel.

The second feature in the model map is low-level 1.809 \MeV\ and 53 GHz 
free-free emission from some open clusters in the Cepheus 
region, in particular from Trumpler 37.
Trumpler 37 is a young, rich, and nearby open cluster that is embedded 
into the bright \HII\ region IC 1396.
Distance estimates based on {\em Hipparcos} data vary between $DM=9.1$ 
and $9.9$ mag (de Zeeuw et al. 1999\nocite{dezeeuw99}; Robichon et al. 
1999\nocite{robichon99}) while our analysis results in a value of
$10.2\pm0.4$ mag.
Using this larger estimate we predict a median 1.809 \MeV\ flux of 
$1.6\times10^{-6}$ \funit.
Adopting the short {\em Hipparcos} distance of \cite{dezeeuw99} would 
result in a flux of $\flxal = 3.9\times10^{-6}$ \funit, making 
Trumpler 37 a potential source of 1.809 \MeV\ \gray\ line 
emission in an area that is little affected by source confusion.
COMPTEL observations (cf. Fig.~\ref{fig:maps}) do not show any feature 
in this area, yet the predicted flux is considerably below the 
sensitivity limit of this instrument.
With its improved sensitivity, the SPI telescope aboard the upcoming
{\em INTEGRAL} observatory may be able to detect this emission, in 
particular if current nucleosynthesis models indeed underestimate \al\ 
production in massive stars (see below).

\subsubsection{Integrated emission}

Finally, the luminosities of all associations have been combined 
by successive marginalisation (Eq.~\ref{eq:total}) to predict the 
total fluxes from the Cygnus region; the results are presented in the 
last row of Tab.~\ref{tab:obresults}.
For the entire Cygnus region we obtain a median 53 GHz flux of 5000 
Jy, close to the observed value of $4200\pm700$ Jy.
The total predicted 1.809 \MeV\ flux, however, is significantly below the 
observations:
while the model predicts a median gamma-ray line flux of
$\flxal = 2.5 \times 10^{-5}$ \funit\ and a $63.8\%$ confidence interval 
of $(1.6 - 4.0) \times 10^{-5}$ \funit, the COMPTEL data suggest a value 
of $(5.8\pm1.5)\times 10^{-5}$ \funit.
Our model predicts only little \fe\ synthesis in Cygnus, and 
consequently, the predicted median 1.137 \MeV\ flux amounts to only
$\flxfe = 2\times10^{-6}$ \funit.
At this level, it seems unlikely that the SPI telescope will detect 
the radioactive decay of the \fe\ isotope in the Cygnus region.

On average, about $80\%$ of the 1.809 \MeV\ emission in our model 
originates from \al\ ejected by stellar winds while only $20\%$ come 
from core-collapse supernovae.
Also, OB associations play a dominant role in the Cygnus region since 
they contribute to almost $80\%$ of the 1.809 \MeV\ emission and to 
about $90\%$ of the free-free luminosity.

Our model predicts $99-172$ O stars and $13-34$ Wolf-Rayet stars in 
the Cygnus region associations, about $\sim25\%$ less than we find in 
our cluster database.
We think that this discrepancy is acceptable, in particular in view of 
the strong variation of the O star number with association age
(Fig. \ref{fig:CygOB2age}).
Note that in any case, the missing O stars cannot explain the apparent 
underestimation of the 1.809 \MeV\ luminosity, since lowering the age 
estimates of some associations (in particular Cyg OB2) to increase the 
number of O stars tends also to lower the \al\ production.
On the other hand, an important increase in the number of O stars will 
also increase proportionally the ionising flux for the Cygnus region, 
leading quickly to an overproduction of Lyc photons.

\section{Discussion}
\label{sec:discussion}

One essential finding from a census of all known clusters and OB
associations and our modelling effort is that we can comfortably reproduce 
the ionising flux from the Cygnus region, while we fall short by a factor of 
2 in explaining \al\ nucleosynthesis.
There are two possibilities that may lead to such a result:
either the nucleosynthesis models we employed underpredict \al\ production 
by a about a factor of 2, or we overpredict the ionising flux by about a 
factor of 2, and in the same time, underestimate the stellar population in 
the Cygnus region by about the same factor.
We tend to believe that the problem is more likely related to the 
nucleosynthesis models, and in the following we will present the 
elements that lead us to this conclusion.

\subsection{Ionisation}
\label{sec:ionisation}

Our prediction of the ionising flux from the Cygnus region relies on 
stellar atmosphere models that predict the stellar Lyman continuum 
luminosities as function of the stellar parameters.
However, the lack of spectroscopic data in the extreme ultraviolet 
shortward of the hydrogen ionisation limit leaves these models 
basically untested in this wavelength region, and we have to consider 
the reliability of the employed atmosphere models in predicting Lyc 
fluxes.

The status of predictions of ionising fluxes has recently been 
reviewed by Schaerer (1998\nocite{schaerer98}, 1999\nocite{schaerer99}).
From a comparison of Lyman continuum photon fluxes predicted by 
various atmospheric models, \cite{vacca96} concluded that they are 
consistent to within $20\%$.
\cite{oey97} compared observed \HII\ region luminosities in the Large 
Magellanic Cloud to the Lyman continuum luminosities predicted from their 
well--determined stellar content, using the CoStar atmosphere models of 
\cite{schaerer97} that we also employed in this work.
They find that on average, \HII\ region luminosities amount to $74\%$ 
of the ionising luminosities predicted by the stellar models, leaving room 
for about $25\%$ flux overprediction by the models.
\cite{hunter90} come to a similar conclusion, using the stellar Lyc 
estimates of \cite{panagia73} applied to small galactic \HII\ regions.
However, accounting for the downward revision of the effective 
temperature scale due to non-LTE line blanketing effects, reduces the 
Lyc fluxes for a given spectral type (Martins et al. 
2002\nocite{martins02}) eliminating the overprediction found by the 
above studies.

On the other hand, it is commonly believed that some fraction of the 
ionising photons escape the \HII\ regions and ionise the diffuse, warm, 
ionised medium (WIM) of the Galaxy (e.g. Oey \& Kennicutt 
1998\nocite{oey98}).
The WIM has been found to comprise $20-53\%$ of the total H$\alpha$ 
luminosity in nearby star-forming galaxies (see Ferguson et al. 
1996\nocite{ferguson96}), a number that is consistent with the 
$\sim25\%$ of ionising photons that are on average missing in the 
\HII\ regions with respect to the model predictions.
Our study is not much affected by the fraction of ionising 
photons escaping into the WIM, since our microwave flux estimate from 
the Cygnus region has been obtained by integrating over a rather larger 
spatial area, including both individual \HII\ regions and the WIM
(see Sect. \ref{sec:observations}).

Alternatively, ionising photons may be also absorbed by dust within 
the \HII\ regions, and a recent study of \cite{inoue01} indicates 
that $\sim20\%$ of the photons may be lost in small galactic \HII\ 
regions in this way.
Using a decomposition of DIRBE far-infrared data, \cite{sodroski97} 
estimated an infrared excess of $1.1-1.7$, corresponding to $5\%-12\%$ 
of Lyc photon absorption within the Galaxy
(Mezger et al. 1974\nocite{mezger74}).
Thus instead of ionising the WIM, the apparent $25\%$ flux overprediction 
may be also (partly) explained by dust absorption.

Whatever the mechanism, $25\%$ ionising flux overprediction by our model 
should be a rather solid upper limit, which is in any case far from 
the required factor of 2 that is needed to explain the \al\ 
nucleosynthesis.

\subsection{Completeness}
\label{sec:complete}

In our aim to predict absolute fluxes from the OB associations and 
young open clusters in Cygnus, we have to be concerned about the completeness 
of our database. 
First, there might be more OB associations in Cygnus than those we 
listed in Table \ref{tab:distance}.
Due to their extent and weak clustering, OB associations are generally 
difficult to identify, in particular if they are nearby.
To estimate the association completeness, we compared the number of O 
stars in the Cygnus area from the SIMBAD database to the number of O 
stars in our stellar database.
Many entries in the SIMBAD database originate from the O star catalogue 
of \cite{garmany82}, which is complete down to a visible magnitude of
$V=10$ mag, corresponding to a maximum $DM+A_{V}=14.3$ mag for O type 
stars.
Comparison of this limit to the distance moduli and reddenings in 
Table \ref{tab:distance} shows that, except for Cyg OB2 and some of the 
heavily reddened or distant open clusters, the SIMBAD data should
indeed be complete for the Cygnus region.

In total we find 171 MK classified O stars in the surveyed area in the 
SIMBAD database while our association database contains 204 O stars.
Our excess in O stars is due to the 120 objects in the heavily reddened 
Cyg OB2 association that are only partially identified in SIMBAD 
(about $50\%$).
Excluding Cyg OB2 from SIMBAD and our database results in 108 O stars
in SIMBAD versus 84 O stars in our database, resulting in a 
completeness of almost $80\%$.
Adding the 120 O stars of Cyg OB2 to both samples increases the O star 
completeness to about $90\%$.
Thus, most known O stars are indeed located in the OB associations of 
our database.

Secondly, there may be young open clusters in Cygnus that escaped 
so far detection due to the heavy obscuration in this area.
Indeed, using {\em 2MASS} near-infrared data, \cite{dutra01} found 21 
new cluster candidates in our survey region ($60\deg < l < 110\deg$), 
which doubles the number of young open clusters in this area.
Also, on basis of {\em 2MASS} data, \cite{duigou02} determined the 
stellar content of 15 new infrared cluster candidates in the area, 
and concluded that they triple the OB population with respect to the 
known open clusters in the centre of the Cygnus X region.
However, compared to the giant association Cyg OB2, they contribute 
only a small fraction, and we find only an additional $10\%$ of OB 
stars in the newly discovered clusters candidates.

Thus, in summary, we believe that our database is fairly complete, 
within about $10-20\%$, and that the 1.809 \MeV\ luminosity puzzle 
can not be explained by a hidden population of massive stars.
In any case, if there would be such a hidden population, it should 
also produce a considerable amount of ionising photons, which would 
lead quickly to an overproduction with respect to the microwave 
observations.

\subsection{Rotation}
\label{sec:rotation}

Throughout this work we have ignored the effects of stellar rotation, 
mainly because there is not sufficient stellar data available for the 
stellar associations we are interested in, and since there are no stellar 
grids available so far that allow for inclusion of rotation in 
evolutionary synthesis calculations.
Rotation firstly affects the colour and luminosity of a star, hence it 
may introduce an artificial age spread in the cluster HRDs
(e.g. de Geus 1990\nocite{degeus90}; Lamers et al. 
1997\nocite{lamers97}).
Since we accounted for age uncertainties in our approach, however, we do 
not believe that neglecting of rotation in the determination of the 
cluster parameters plays an important role.

\begin{figure}[tb]
  \includegraphics[width=8.8cm]{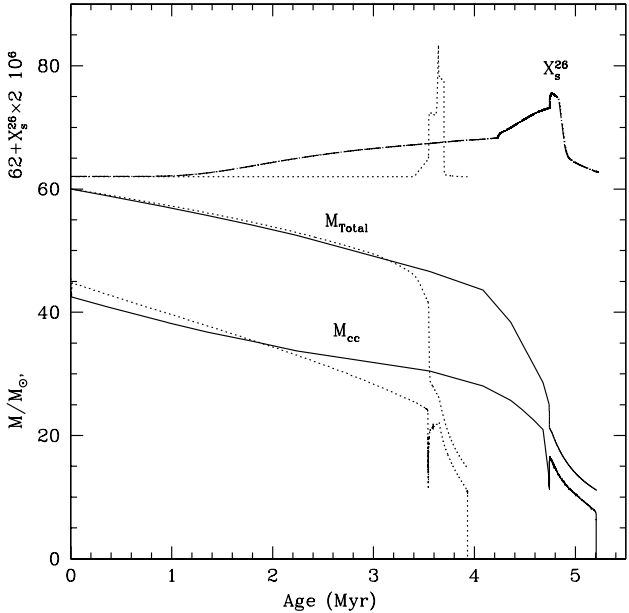}
  \caption{Evolution as a function of time of the total mass of the star,
           M$_{\rm Total}$, of the mass of the convective core, M$_{\rm cc}$ 
           and of the surface abundance of \al\ in mass fraction, 
           ${\rm X}^{26}_{\rm s}$, for a rotating (continuous lines) and a 
           non--rotating (dotted lines) 60 \Msol\ stellar model
           at solar metallicity. The initial equatorial velocity for the 
           rotating model is 300 km s$^{-1}$.}
  \label{kipwral}
\end{figure}

\begin{figure}[tb]
  \includegraphics[width=8.8cm]{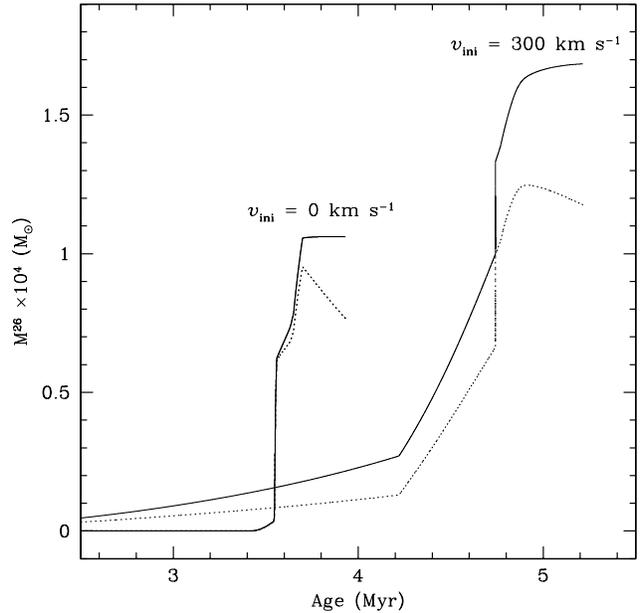}
  \caption{Continuous lines show the evolution as a function of time of the
           integrated mass of \al\ ejected by stellar winds for a non-rotating 
           and a rotating 60 \Msol\ stellar model at solar metallicity. The 
           initial equatorial velocities are indicated. The dotted lines show 
           the evolution of the mass of non-decayed \al.}
  \label{venwral}
\end{figure}

Much more important may be the impact of rotation on stellar evolution 
and in particular the nucleosynthesis of \al.
To illustrate this point, we show in Fig.~\ref{kipwral} the evolution
of the surface \al\ abundance as a function of time for two 60 \Msol\ stellar 
models at solar metallicity with and without rotation (Ringger 2000\nocite{ri00}). 
The models were computed with the same physical ingredients as used in 
\cite{MMV}.
Let us note in particular that the mass-loss rates used by 
\cite{ri00} are smaller by a factor of $2-3$ than the mass-loss rates 
used in the stellar models that have been applied to calculate the 
\al\ yields in this work.

In the case of the rotating models, surface enrichments begin to occur at a 
much earlier stage than in the non-rotating model (see Fig.~\ref{kipwral}).
Indeed rotational mixing brings freshly synthesised \al\ at the surface
well before shells, having experienced CNO processing, are uncovered by
the stellar winds. 
This explains why the \al\ abundance at the surface of the rotating model
increases in a much smoother way than at the surface of the non-rotating
model.
One can see also that rotational mixing slows down the decrease in mass of the 
convective core and increases the Main Sequence lifetime. 
This results from the diffusion of hydrogen into the convective core. 
Some $^{25}$Mg will also migrate from the radiative envelope into the 
convective core where it will be transformed into \al. 
These effects tend to increase the quantity of \al\ ejected by the stellar 
winds. 

Fig.~\ref{venwral} shows the evolution of ${\rm M}^{26}$, the integrated mass
of \al\ ejected by the stellar winds by the rotating and non-rotating 60
\Msol\ models. 
This quantity is obtained by estimating the integral 
\begin{equation}
{\rm M}^{26}(t)=\int_0^t X^{26}_s(t') \dot M(t') {\rm d}t',
\end{equation}
where $\dot M$ is the mass loss rate. 
The luminosity at 1.809 \MeV\ is, at each time, proportional to the mass of 
non-decayed \al\ (see the dotted lines in Fig.~\ref{venwral}).
One sees that the maximum luminosity reached by the rotating model is shifted 
towards later time and is only slightly increased with respect to the 
non-rotating model. 
This occurs because in the rotating model,  the release of the \al\ is 
distributed over a much longer time, and therefore, the radionuclide has 
more time to decay.
Thus even if the quantity of \al\ released by the rotating model is 
significantly greater (for this particular model the enhancement factor is 
$\sim$1.6), the enhancement of the luminosity is not as great.
The above examples show that rotation, all other physical ingredients 
being kept equal, tends to increase the total quantity of \al\ 
ejected, slightly increases the maximum 1.809 \MeV\ luminosity, and 
shifts the ejection of the bulk of \al\ to later times.

How the results of the present population synthesis models would be affected
by rotation?
First of all, let us recall that the WR stellar models used in the present 
population synthesis models implicitly account in some way for non-standard 
physical mechanisms. 
Indeed these models were computed with artificially enhanced mass loss rates 
which enabled to account in a satisfactorily manner for the variation with 
the metallicity of the number of WR to O-type stars 
(Maeder \& Meynet 1994\nocite{maeder94}).
Thus the reproduction of this observational constraints in some way calibrates
these stellar models.
If we compare the total quantity of \al\ ejected by our rotating 
60 \Msol\ model with the non-rotating 60 \Msol\ model with enhanced mass loss 
rates, used in the present population synthesis model, we find a modest 
\al\ yield increase of 13\%.

Does this mean that rotation has to be discarded as a possible mechanism for
solving the \al\ deficit discussed in the present paper?
We think it is too early to reach such a conclusion. 
Since the work of \cite{ri00}, new improvements of the physics of rotation 
have been achieved especially concerning the shear diffusion coefficient and 
the impact of rotation on the mass loss rates 
(Maeder \& Meynet 2000\nocite{maeder00}, 2001\nocite{maeder01}).
Predictions of precise enhancement factors requires the computation
of stellar models for different initial masses and rotational velocities. 
At the present time, no such grids of models, predicting the 1.809 \MeV\ 
luminosity, have been computed.
We therefore defer a more quantitatively assessment of rotation on gamma-ray 
line emission in Cygnus to future work.

\subsection{Binarity}
\label{sec:binarity}

Binarity may be another mechanism that could alter the \al\ yields from 
massive stars, yet we did not account for their effects due to the 
involved complexity and poorly known details.
In particular, only close binary systems are concerned, and there is 
not sufficient data available for the associations we are interested in to 
assess their number and mass distribution.

Two effects may impact the nucleosynthesis in close binary systems:
tidal interactions in close binary systems, and mass transfer by Roche 
Lobe overflow.
Tidal effects are expected to deform the star and therefore induce 
instabilities reminiscent of those induced by rotation.
In particular, tidal forces and orbital motions will induce rotation even 
in initially non-rotating stars.
To our knowledge, this effect has never been studied in 
nucleosynthesis calculations, despite its potentially important 
consequences, as for instance by homogenising the stars.

If mass transfer occurs, the removal of part of the envelope of the 
donor star may favour the appearance of \al\ on the surface, in a 
similar fashion as mass--loss acts through stellar winds.
More important changes may occur for the gainer in systems with 
$M < 40\Msol$, as e.g. shown by the preliminary studies of 
\cite{braun95} and \cite{langer98}.
The latter suggest a scenario in which mass transfer onto the 
secondary leads to a rejuvenation which alters its subsequent evolution.
They predict an increase by $2-3$ orders of magnitudes of the 
hydrostatically produced \al\ yield due to a reduction of the delay 
between production and ejection of \al, which may possibly enhance 
the total \al\ production from Type II supernovae by a factor of 
about 2 (Langer, priv. communication).
Since, however, supernovae are very rare in Cygnus, it is highly 
doubtful that such a mechanism could solve the \al\ yield puzzle.

\section{Conclusions}
\label{sec:conclusions}

\subsection{Nucleosynthesis}
\label{sec:nucleosynthesis}

Our modelling effort of the gamma-ray line emission from OB associations 
and young open clusters in the Cygnus region has revealed a possible 
shortcoming of actual nucleosynthesis models in explaining \al\ 
production.
We find a 1.809 \MeV\ flux underestimation of about a factor of 2 that is 
difficult to explain by other means than a modification of current 
nucleosynthesis models for single, non-rotating stars.
From preliminary calculations, it appears difficult for rotation alone 
to significantly enhance \al\ production, although this conclusion 
needs to be checked by more detailed calculations in the future.

Only little \fe\ production is predicted in Cygnus by our model, 
mainly related to the low number of recent supernova events in this 
region.
A detection of the 1.137 \MeV\ and 1.332 \MeV\ lines from the 
radioactive decay of \fe\ by {\em INTEGRAL} would therefore present a 
big surprise.
In the case of such a detection, \fe\ production by hydrostatic 
helium burning in Wolf-Rayet stars -- which we have not included in 
our model due to the low yield predictions -- should then be seriously 
reconsidered (Arnould et al. 1997\nocite{arnould97}).

\subsection{Other gamma-ray signatures}
\label{sec:signatures}

In addition to \al\ and \fe, massive star associations may produce 
further radioactive isotopes during the supernova explosion of massive 
stars, such as \co\ and \ti, yet their short lifetimes of $112$ days 
and $87$ years, respectively, make their observation impossible in 
absence of a very recent event.
However, \co\ and \ti\ decay under positron emission (like \al), and the 
annihilation of positrons with electrons of the interstellar medium on 
time scales of a few $10^5$ years may provide a reverberation of the 
short-lived, extinct, radioactivities.
Hence, the observation of the 511 keV positron annihilation line may 
provide an independent measure of the supernova activity in massive 
star associations (and in particular in Cygnus), although the 
interpretation of the observations will be complicated by the 
annihilation physics, the positron transport, and the positron escape 
fraction from the expanding supernova remnants.

\subsection{Super star clusters}
\label{sec:superstarclusters}

Gamma-ray as well as free-free emission in the Cygnus region seems to 
be dominated by a single, extremely massive association: Cyg OB2.
Indeed, Cyg OB2 is merely a prototype of a young globular cluster 
than an OB association (Kn\"odlseder 2000\nocite{knoedl00}).
There are examples of further super star clusters in the Galaxy, such as
NGC 3603 (Moffat et al. 1994\nocite{moffat94}),
the Arches and Quintuplet clusters near the galactic centre 
(Figer et al. 1999\nocite{figer99}),
or the W49A cluster (Conti \& Blum 2002\nocite{conti02}).
Most of these super star clusters are partially or totally obscured in 
the visible by the absorbing effects of intervening and/or local 
interstellar dust, and it is unclear how many of them exist 
throughout the entire Galaxy.
Making the simplifying assumption that super star clusters are 
distributed uniformly throughout the Galaxy within a galactocentric 
distance of 15 kpc, and taking that Cyg OB2 is apparently the only such 
object within 1.5 kpc, one may expect 100 super star clusters in our 
Galaxy.
Assuming that they all produce $10^{-2}$ \Msol\ of \al, similar to 
our Cyg OB2 model prediction, a total \al\ production of 1 \Msol\ is 
expected from these objects.
Taking into account the \al\ yield underestimation of a factor of 2 
brings this mass to 2 \Msol, comparable to the observed galactic \al\ 
mass of $2-3$ \Msol\ (Diehl et al. 1995\nocite{diehl95}; 
Kn\"odlseder 1999\nocite{knoedl99}).
Thus, a considerable fraction of \al\ maybe indeed produced by such 
super star clusters, and 1.809 \MeV\ gamma-ray observations with 
sufficient angular resolution and sensitivity may help to uncover and 
to study them throughout the Milky-Way.

\subsection{The Cygnus superbubble}

Our age determination of clusters in the Cygnus region sheds some doubt 
on the triggered star formation scenario that has been proposed to explain 
anomalous stellar proper motions in the area
(Comer\'on \& Torra 1994\nocite{comeron94}).
In this scenario, the central association Cyg OB2 is supposed of 
having formed a shell blown by stellar winds and supernovae, the 
Cygnus superbubble, that subsequently gave birth to the surrounding 
associations Cyg OB1, OB3, OB7 and OB9 due to gravitational shell instability.
However, our age estimate for Cyg OB2 is inferior to that of the 
surrounding associations, making triggered star formation in this area
unlikely.
In contrast, the anomalous stellar proper motions are equally well 
explained by supposing that they result from dynamically ejected 
runaway stars from Cyg OB2 (Comer\'on et al. 1998\nocite{comeron98}), 
in particular since the proposed expansion age of $\sim4$ Myr is 
compatible with our age estimate for Cyg OB2.

\subsection{Cygnus and the Galaxy}
\label{sec:galaxy}

Finally, we want to mention that not only the Cygnus region but also 
the Galaxy as a whole suffers probably from an underproduction of \al\ 
when current, non-rotating nucleosynthesis models are considered.
By combining the same nucleosynthesis models that have been used in 
this work with the atmosphere models of \cite{schaerer97}, and 
normalisation to the observed galactic Lyman continuum luminosity,
\cite{knoedl99} predicted a galactic \al\ mass of $1.6$ \Msol, significantly 
below the observed value of $2-3$ \Msol.
Only the inclusion of the galactic metallicity gradient, which 
considerably enhances \al\ ejection through stellar winds in the 
Wolf-Rayet phase towards the galactic centre, provides a sufficient increase 
of the galactic \al\ production, bringing the model predictions in better 
agreement with the observations. 

Yet, metallicity is of no help in the Cygnus region, where even 
slightly subsolar abundances are reported 
(Daflon et al. 2001\nocite{daflon01}).
On the other hand, if another process, such as rotation, is needed to 
explain the \al\ production in Cygnus, it should also be at work for 
the entire Galaxy, leading to a potential \al\ overproduction if 
combined with the metallicity enhancement.
In fact, if \al\ production by massive stars would indeed follow the
$(Z/Z_\odot)^{2}$ dependence predicted by non-rotating stellar 
models, a break-down of the tight correlation between 1.809 \MeV\ and 
galactic free-free emission would be expected due to the inverse metallicity 
dependencies of both emission processes -- the 1.809 \MeV\ intensity should 
be enhanced in the metal-rich inner regions of the Galaxy due to enhanced 
\al\ production, while the free-free emission should be reduced due to lower 
electron temperatures (see Kn\"odlseder 1999\nocite{knoedl99}).

The existing data show no indication for such an anti-correlation.
However, it is likely that rotation reduces the metallicity dependence of
Wolf-Rayet star yields.
The fact that at low metallicity, there is less $^{25}$Mg, might be somewhat
compensated by the fact that when the metallicity decreases, the mixing is 
more efficient, bringing more $^{25}$Mg from the radiative envelope
into the core and more \al\ from the convective core into the radiative
envelope. 
The mixing efficiency increases when the metallicity decreases because, when 
$Z$ decreases, the stars are more compact, the internal gradients of the 
angular velocity are steeper and the stars loose less angular momentum by 
mass loss through stellar winds (see Maeder \& Meynet 2001\nocite{maeder01}).
Secondly, when rotation is accounted for, the mass loss rate plays a less
important role in the WR formation process (see Fig.~\ref{kipwral}).
Thirdly, at low metallicity,  stars loose less angular momentum. 
As a consequence they can more easily reach the break-up limit at a given 
stage during their evolution. 
Very high mass loss rates ensue even if the metallicity is low. 
Finally, there are some indirect indications, that the proportion of fast
rotators increases when the metallicity decreases 
(Maeder et al. 1999\nocite{mgm}). 

Since rotation seems to erode the metallicity dependence of 
Wolf-Rayet star yields, it provides an appealing mechanism to explain 
the absence of a 1.809 \MeV\ -- free-free emission anti-correlation.
Remains to be seen if rotation can also be a solution to the \al\ 
yield puzzle -- both for the Cygnus region and the Galaxy as a whole.

\begin{acknowledgements}
This research has made use of the SIMBAD database, operated at CDS, Strasbourg, 
France and the WEBDA database, compiled by Jean-Claude Mermilliod,
Institute of Astronomy of the University of Lausanne, Switzerland.
\end{acknowledgements}

\appendix
\section{OB associations}

\subsection{Cep OB1}

Our data are based on the work of \cite{garmany92} complemented by early 
type stars within the field of the association that we extracted from the 
SIMBAD database.
The large angular extent of $15\deg \times 4\deg$ together with a 
distance of $3.6$ kpc implies linear dimensions of $950 \times 250$ 
pc, considerably larger than the typical size of OB associations 
(Garmany 1994\nocite{garmany94}).
Already \cite{moffat71} noted for this reason that Cep OB1 may be the 
combination of several OB associations.
Indeed, the HRD suggests the existence of two subgroups of different age 
that we term Cep OB1a ($2-5$ Myr) and Cep OB1b ($9-18$ Myr).
There are 4 WR stars in the field of Cep OB1 which are possibly 
associated to the young subgroup.
However, the subgroups do not separate spatially into two distinct 
groups but are rather immersed into each other.
Thus the large size of Cep OB1 still remains unexplained, and it is 
possible that Cep OB1 splits either into even more subgroups or that 
the stars are physically unrelated.

\subsection{Cep OB2}

Cep OB2 has been recently revisited by \cite{dezeeuw99} by means of {\em Hipparcos} 
data, and we based our analysis on their list of probable members.
We complemented the {\em Hipparcos} data by spectroscopic information from 
the SIMBAD database, and carefully excluded possible members of the 
cluster Trumpler 37 from the dataset.
The resulting HRD shows a relatively nice main sequence, and our 
distance modulus of $9.1\pm0.3$ mag agrees well with the {\em Hipparcos} value 
of $8.9$ mag (de Zeeuw et al.~1999\nocite{dezeeuw99}).
However, in contrast to \cite{dezeeuw99}, we do not confirm a physical 
relation between the association and the open clusters NGC 7160 and 
Trumpler 37 since both lie at significantly larger distances.

\subsection{Cyg OB1}

The Cyg OB1 association spatially overlaps with the young open clusters
Berkeley 86, Berkeley 87, IC 4996, and NGC 6913, and we will argue below 
that all four clusters are likely physically related to the association.
We based our stellar census on the compilations of \cite{garmany92} 
and \cite{zakirov99} which include also member stars from the open clusters.
Stars that coincide spatially with one of these clusters have 
therefore been excluded and put in the respective cluster database.
Conversely, stars from the cluster databases that are located outside 
the classical cluster boundaries were included in the Cyg OB1 list.
We added spectroscopic information from the SIMBAD database and added 
also the Wolf-Rayet stars that are likely associated to Cyg OB1 
(van der Hucht 2001\nocite{hucht01}).
The resulting HRD shows a clear though slightly broad main sequence, 
and the relatively flat IMF slope of $\Gamma=-1.0$ suggests that our 
database is probable not complete for low mass stars.

\subsection{Cyg OB2}

Cyg OB2 is one of the most massive stellar associations known in our 
Galaxy and houses about $\sim120$ O type stars 
(Kn\"odlseder 2000\nocite{knoedl00}).
We based our analysis on the spectrophotometric survey of \cite{massey91} 
although we recognise that their observations cover only the central 
part of the association area.
We therefore adopt the results from \cite{knoedl00} for the total mass 
normalisation, yet rely on the more precise spectroscopic information 
from \cite{massey91} for the determination of the remaining 
association parameters.
We added the three Wolf-Rayet stars WR144, 145, and 146 that coincide 
spatially with Cyg OB2 to the database, although their physical 
relation to the association is not established 
(van der Hucht 2001\nocite{hucht01}).
The resulting HRD shows a nice and clearly defined main sequence 
(cf.~Fig.~\ref{fig:CygOB2}).
Our association parameters agree well with those found in other 
studies (e.g.~Massey \& Thompson 1991\nocite{massey91}, 
Torres-Dodgen et al.~1991\nocite{torres91}).

\subsection{Cyg OB3}

For Cyg OB3 our analysis is based on the compilation of \cite{garmany92} 
complemented by spectroscopic data from the SIMBAD database.
On basis of spatial location possible members of the open clusters 
NGC 6871 and Biurakan 2 were excluded, and inversely, possible 
non-members of both clusters were included in the Cyg OB3 database.
We also added the Wolf-Rayet stars WR134 and 135 
(van der Hucht 2001\nocite{hucht01}).
The resulting HRD is rather sparse yet shows a moderately well 
defined main sequence which, however, lacks low mass association 
stars.
As consequence we obtain formally an extremely flat IMF slope of 
$\Gamma=-0.3\pm0.4$ which is probably heavily biased by the 
incompleteness of our database for faint stars (this should however 
not affect our nucleosynthesis predictions since only massive stars 
will contribute significantly to the yields).

\subsection{Cyg OB7}

The data for this extremely sparse association are again based on the 
list of \cite{garmany92} complemented by spectroscopic information 
from the SIMBAD database.
Based on an analysis of Hipparcos data \cite{dezeeuw99} argues that 
Cyg OB7 is likely a chance projection of massive stars since no common 
motion has been detected for the members (the authors draw the same 
conclusion for Cyg OB4).
Our HRD shows a very sparsely populated yet reasonably narrow main 
sequence, therefore we kept the association in our list of massive star 
populations.

\subsection{Cyg OB8}

The data for this association were compiled from \cite{humphreys78} and 
the SIMBAD database.
The HRD is extremely sparse and no clear main sequence can be 
identified, making the physical reality of this association extremely 
doubtful.
We reflect this uncertainty by attributing a large age uncertainty 
($1-14$ Myr) to Cyg OB8.

\subsection{Cyg OB9}

Our database was compiled from the list of \cite{garmany92} complemented 
by spectroscopic information from SIMBAD.
Possible members of NGC 6910 have been excluded from the list based 
on spatial location.
The HRD of this association is rather sparse, yet a broad main sequence is 
perceptible which indicates an age spread of $2-5$ Myr.

\subsection{Lac OB1}

The nearby association Lac OB1 has been recently studied by \cite{dezeeuw99} 
using {\em Hipparcos} data, and we based our analysis on their list of likely 
members, complemented by SIMBAD data.
Lac OB1 shows a reasonably well defined main sequence with a turn-off 
age around $12-15$ Myr.
There is one star deviating from the main sequence at the bright end 
(10 Lac) and it is questionable if this star is physically related to 
the association (the {\em Hipparcos} parallax places this star at the 
near side of the association).
The most intriguing discrepancy with respect to the study of \cite{dezeeuw99} 
is the distance estimate towards the association.
The {\em Hipparcos} parallax measurement of $368\pm17$ pc translates 
into a distance modulus of $7.8\pm0.1$ mag while we obtained a 
significantly larger value of $9.0\pm0.5$ mag.
De Zeeuw et al.~(1999)\nocite{dezeeuw99} pointed already out that their 
distance estimate is significantly smaller than most spectrophotometric 
estimates, but attributed this discrepancy to the significant modification 
of the list of members in their work.
We cannot confirm this hypothesis since we used the same member list 
as \cite{dezeeuw99} and still find a significantly larger distance 
estimate for Lac OB1.

A possible explanation of this discrepancy may come from the fact 
that most MK classified stars in Lac OB1 are of spectral type B2-3.
In this spectral domain the absolute visual magnitude $M_{V}$ varies 
extremely rapidly with spectral type (see e.g.~Fig.~1a of de Geus 
1990\nocite{degeus90}) and consequently the absolute luminosity of 
these stars is only poorly defined.
Comparison of our Lac OB1 data to the {\em Hipparcos} parallaxes suggests 
that absolute magnitude estimates for B2-3 V stars are about $1.4$ 
mag too bright.
It is not clear to us if this overestimation is due to a spectral 
misclassification of the stars (which seems unlikely due to the high 
number of stars involved), due to the insufficiency of the stellar 
calibrations (we used the calibrations of Humphreys \& McElroy 
1984\nocite{humphreys84} and Schmidt-Kaler 
1982\nocite{schmidtkaler82}), or due to a subtle effect of stellar 
rotation (e.g.~Lamers et al.~1997\nocite{lamers97}; however we did not 
find any dependence of luminosity overestimation on rotational 
velocity).
We note, however, that it appears as a general feature in the analysis 
of our stellar populations that stars in this spectral domain tend to be 
too bright (see for example Fig.~\ref{fig:CygOB2} which shows the same 
trend for stars with $\lTeff \sim 4.3$ in Cyg OB2).
It is however not within the scope of the present paper to revise the 
absolute magnitude calibration for early B-type stars, but it is 
obvious that there is a pressing need for a detailed look at 
this problem (e.g.~Massey et al.~1995\nocite{massey95}).

\subsection{Vul OB1}

For Vul OB1 our database is based on the compilation of \cite{garmany92}, 
complemented by spectroscopic information from SIMBAD.
From their spatial location possible members of NGC 6823 have been 
excluded from the list.
The HRD of this association is rather sparse, and no clear main sequence 
is perceptible.
The physical reality of Vul OB1 is therefore highly questionable.

\section{Open clusters}

\subsection{Berkeley 86}

Berkeley 86 has been proposed by \cite{blaha89} as one of the three 
nuclei of Cyg OB1, the others being NGC 6913 and IC 4996.
The WEBDA data on Berkeley 86 are contaminated by stars of the Cyg 
OB1 association and the cluster IC 4996, hence a careful separation 
of these stars has been conducted on basis of stellar positions.
The resulting cluster parameters are compatible with the values found
in the literature (e.g.~Deeg \& Ninkov 1996\nocite{deeg96}).
The distance modulus, reddening, and age of Berkeley 86 is in agreement 
with the parameters of Cyg OB1, supporting its physical 
connection to the association.
There is one Wolf-Rayet star in the field of Berkeley 86 (WR139), an 
eclipsing binary system of type WN5 - O6III-V 
(van der Hucht 2001\nocite{hucht01}).
Our age estimate of $3-5$ Myr is compatible with the physical 
association of the system to the cluster.

\subsection{Berkeley 87}

For Berkeley 87, the data in the WEBDA database are essentially based 
on the UBV photometric survey of \cite{turner82} and the spectroscopic 
observations of \cite{massey01}.
From the former data the authors determined a distance modulus of $9.88$ 
mag for the cluster, while the new spectroscopic data of \cite{massey01} 
suggest a high value of $11.0$ mag.
We also find a high distance modulus of $11.4$ mag for the combination of 
all available data, which places Berkeley 87 at the same distance as the Cyg 
OB1 association.
The spatial overlap with Cyg OB1 makes it highly likely that Berkeley 87 is 
indeed physically connected to this association.
Among all clusters that we examined, Berkeley 87 shows the highest reddening 
($E(B-V)=1.63$), reflecting the heavy obscuration by a large molecular 
cloud complex in this area.
Berkeley 87 houses one of the rare type WO2 Wolf-Rayet stars (WR142) 
near its centre (van der Hucht 2001\nocite{hucht01}), and our age estimate 
of $3-6$ Myr seems compatible with the existence of such an object in the 
cluster.

\subsection{Berkeley 94 and 96}

Only photometric data are available for the two clusters from which we 
infer a distance modulus of $13.6$ mag for both.

\subsection{Biurakan 2}

For Biurakan 2, we supplemented the WEBDA data with spectroscopic 
information from the SIMBAD database.
We removed stars number 126 and 131 from the list since their location 
in the $E(B-V)$ versus $DM$ diagram clearly identifies them as foreground 
objects.
Our resulting distance modulus of $11.8$ mag is in excess to other works 
(e.g.~Dupuy \& Zukauskas 1976\nocite{dupuy76}), which is mainly 
explained by the removement of stars number 126 and 131.
We note that Biurakan 2 lies at the edge of the Cyg OB3 and its distance 
modulus and reddening are compatible with the parameters of this association.
Yet, our age estimate of $24-30$ Myr makes their physical relation 
questionable.

\subsection{IC 4996}

IC 4996 is situated in the field of Cyg OB1 and has been suggested as 
nuclei of the association (Garmany \& Stencel 1992\nocite{garmany92}, 
and references therein).
Our cluster parameters corroborate this hypothesis.

\subsection{IC 5146}

The cluster IC 5146 is related to a spherical emission-reflection 
nebula which is excited by the most massive star in the field, the
B1V star BD+46\deg 3474 (e.g.~Wilking et al.~1984\nocite{wilking84}). 
The WEBDA database for IC 5146 is heavily contaminated by fore- and 
background stars, and severe boundaries on $E(B-V)$ were necessary to 
extract the cluster stars.
The resulting HRD is very sparsely populated, with 
BD+46\deg 3474 being the only massive object.
Indeed, for this star we derive a distance modulus of $9.2$ mag while 
the average cluster distance amounts to $10.3$ mag.
Excluding BD+46\deg 3474 results in an even larger DM of $10.8\pm0.3$ 
mag, suggesting that BD+46\deg 3474 may indeed lie in front of the 
cluster.
In this case, the earliest spectral type in IC 5146 would be as late 
as B8V, making it too old for our purpose.
We therefore excluded the cluster from the analysis.

\subsection{NGC 6823}

NGC 6823 is an extremely rich cluster for which abundant photometric and 
spectroscopic information is available.
The colour-colour diagram of the database shows a very broad main sequence 
between $0.5 < E(B-V) < 1.2$, illustrating the variable extinction in this 
area.
Indeed, NGC 6823 lies in the middle of the moderately bright \HII\ 
region NGC 6820, and variable extinction is common in such a 
configuration.
NGC 6823 is proposed as the nucleus of the Vul OB1 association, and 
indeed our cluster parameters agree well with that of the association.

\subsection{NGC 6871}

NGC 6871 has been suggested as nucleus of the Cyg OB3 association 
(e.g.~Garmany \& Stencel, 1992\nocite{garmany92}) and the WEBDA 
database for this cluster contains a substantial number of 
association members but also field stars.
From their spatial and evolutionary parameters, roughly 20 stars were excluded 
from the WEBDA list and included in the Cyg OB3 database, yet we admit 
that this membership assignment is a rather arbitrary process.
Nevertheless, Cyg OB3 and NGC 6871 have very similar distances and 
ages, and we could probably have treated both as a single object.
The WEBDA data have been complemented with spectral types from \cite{massey95} 
and from the SIMBAD database.
Near the centre of NGC 6871 we find the type WN5 Wolf-Rayet star WR133 
(van der Hucht 2001\nocite{hucht01}).
Our age estimate of $5-6$ Myr is compatible with the presence of such 
an object in NGC 6871.

\subsection{NGC 6883}

The WEBDA database contains only UBV magnitudes for 9 stars in this 
cluster resulting in a sparsely populated main sequence.
The earliest spectral type in the cluster is B1III.
NGC 6883 lies at the edge of Cyg OB3 and the cluster reddening is 
similar to that of the association.
Yet the cluster is marginally closer than Cyg OB3, and in particular 
the age estimate is not compatible with a contemporaneous formation of 
both stellar groups.
We therefore consider a physical relation as doubtful.

\subsection{NGC 6910}

NGC 6910 lies at the edge of the Cyg OB9 association, yet their physical 
relation is questioned by their significantly different distance moduli 
(see also Garmany \& Stencel 1992\nocite{garmany92}).
We removed a small fraction of the stars in the WEBDA database due to 
their particular low $E(B-V)$ values, and updated spectral types from 
the SIMBAD database.

\subsection{NGC 6913}

This young open cluster, also known as M29, is situated in the 
field of Cyg OB1 and is considered as one of the nuclei of this 
association.
\cite{wang00} recently obtained spectroscopic observations of 100 
probable cluster members from which they derived a distance modulus of 
10.17 mag, much closer than the estimate of \cite{massey95} of 
$11.71$ mag based on spectroscopic data.
We examined both the \cite{massey95} and \cite{wang00} datasets and 
found that the latter is probably biased by pre-main sequence and 
foreground stars.
As noted by \cite{wang00}, many stars in their sample show an abnormal 
reddening slope, hence we excluded all objects with $q_{r}$ outside the 
interval $0.5-1.5$ from the dataset.
Already this simple selection increases the distance modulus of the 
\cite{wang00} dataset to $11.1$ mag.
Figure 6 in \cite{wang00} shows that only stars with 
$0.9 < E(B-V) < 1.2$ show a spatial concentration to the cluster centre.
Using the more relaxed condition $0.6 < E(B-V) < 1.5$ we obtain a
distance modulus of $11.3\pm1.0$ mag, while the more severe restriction
$0.9 < E(B-V) < 1.2$ results in $11.7\pm0.6$ mag. 
We therefore adopt a distance of $11.3\pm1.0$ mag for NGC 6913 which 
places the cluster at the same distance as Cyg OB1, strengthening the 
hypothesis of being a nuclei of this association.

\subsection{NGC 7067}

For this poor cluster spectroscopic information is only available 
for three stars (numbers 2, 4, and 5) in the WEBDA database.
However, our analysis suggests that these three stars are all 
foreground objects that are superimposed on the cluster area.
Excluding these stars results in a rather large distance modulus of 
14.2 mag.

\subsection{NGC 7128}

NGC 7128 has been recently studied by \cite{balog01} who determined a 
distance modulus of $13.0\pm0.2$ mag and an age above 10 Myr.
The WEBDA database for this cluster contains spectroscopic 
information for four stars which again are considerable closer than 
the remaining cluster stars.
Thus, we excluded these four objects from the analysis which resulted 
in a distance modulus and reddening that is agreement with the analysis of
\cite{balog01}.

\subsection{NGC 7160}

The WEBDA database contains photometric and spectroscopic information 
for this cluster.
We exclude stars number 407 and 487 from the list since they are too far 
from the cluster centre.
NGC 7160 has been discussed as possible nuclei of Cep OB2 
(Garmany \& Stencel 1992\nocite{garmany92}), but our 
distance modulus estimate places the cluster behind this association.

\subsection{NGC 7235}

The WEBDA database for this cluster is heavily contaminated by 
non-members which we tried to exclude by applying reddening 
selections.
The very steep IMF slope of $\Gamma=-2.6$ indicates that we probably 
did not fully succeed in eliminating faint non-member stars from the 
database.
NGC 7235 lies at the edge of the Cep OB1 association and indeed our distance 
modulus estimates for both objects are fairly compatible.
Our age estimate of $4-5$ Myr suggests a possible physical relation to 
the young subgroup of the association (Cep OB1a).

\subsection{NGC 7261}

NGC 7261 is another cluster at the edge of the Cep OB1 association, 
and our distance estimate again suggests that both objects could be 
physically related.
Our age estimate of $12-20$ Myr suggests that NGC 7261 is related to 
the old subgroup of the association (Cep OB1b).

\subsection{NGC 7380}

The young open cluster NGC 7380 is associated with a bright \HII\ region, 
and as often in such a configuration, there is a large scatter in extinction 
of the cluster stars, ranging from $0.5 < E(B-V) < 0.9$.
We excluded a few stars from the WEBDA database since they were too 
far from the cluster centre and/or their reddenings did not correspond 
to the reddenings found in their vicinity.
A reddening selection was applied to reject further non-members.
NGC 7380 also lies in the field of Cep OB1, and all cluster parameters 
(reddening, distance, age) are compatible with a physical relation to 
the young subgroup (Cep OB1a).

\subsection{Roslund 4}

Roslund 4 is a small open cluster for which the WEBDA database 
contains 12 objects (stars number 13 and 14 have been removed since 
they are local photoelectrical standards and are not part of the 
cluster).
Our distance estimate is identical to that of \cite{racine69}, which 
is no surprise since we used the same data.
No spectroscopic information is available for the stars in Roslund 4, 
making an age estimate rather difficult.
From the luminosity and temperature of the most massive stars in the 
cluster ($\sim10$ \Msol) we infer a highly uncertain age in the range 
$5-15$ Myr.

\subsection{Roslund 5}

The WEBDA data have been complemented by spectral types from the 
SIMBAD database and membership data from \cite{baumgardt98}.
Our distance estimate of $8.6\pm0.3$ agrees nicely with the {\em 
Hipparcos} value of $8.6\pm0.4$ (Baumgardt 1998\nocite{baumgardt98}).
The earliest star in the cluster is of spectral type B3V and based on 
the location of this star in the HRD we estimate an age between 
$20-30$ Myr for this cluster.

\subsection{Ruprecht 175}

The WEBDA database for Ruprecht 175 is based one the work of 
\cite{turner98} and contains, as the author states, a large fraction 
of probable non-member stars.
Based on his Table 5, we selected probable cluster members from the 
database, resulting in a distance modulus of $11.3$ mag for Ruprecht 175.
In contrast, \cite{turner98} obtained a distance modulus of only 
$10.4$ mag.
Our distance estimate is based on the 3 stars with spectroscopic 
information in the database which all show a distance modulus around 
$11.3$ mag.
\cite{turner98} added also photometric data which systematically show 
smaller distance moduli, pushing the mean cluster distance to a lower 
value.
Due to the high uncertainties involved in photometric distance 
estimates, we prefer the spectroscopic distance, although it is only 
based on a small sample of stars.
The earliest star in Ruprecht 175 is of spectral type B3V and based on 
the location of this star in the HRD we estimate an age between 
$20-30$ Myr for this cluster.

\subsection{Trumpler 37}

Trumpler 37 is embedded in the centre of the bright \HII\ region IC 
1396, and is suggested to form the nucleus of the Cep OB2 association.
The cluster has been studied by numerous authors, and consequently, 
there is a rich database of stellar data available.
The WEBDA database contains a substantial fraction of non-members 
which we exclude by means of reddening selections.
Our distance modulus of $10.2\pm0.4$ mag is consistent with other works 
(e.g.~Garrison \& Kormendy 1976\nocite{garrison76}) and agrees in 
particular with the {\em Hipparcos} estimate of $9.9\pm0.8$ mag 
(Robichon et al.~1999\nocite{robichon99}).
This distance is significantly larger than our estimate of $9.1\pm0.3$ mag 
for Cep OB2, rejecting claims that Trumpler 37 is physically related 
to this association (e.g.~de Zeeuw et al.~1999\nocite{dezeeuw99}).
Only an abnormally high ratio of absolute-to-selective absorption of
$R_{V} \simeq 5$ would bring Trumpler 37 to a distance that is 
compatible with that of Cep OB2, yet a variable extinction analysis of 
the cluster stars results in $R_{V} = 3.0 \pm 0.3$, compatible with 
our assumption of $R_{V} = 3.1$ (see also Morbidelli et al. 
1997\nocite{morbidelli97}).
We therefore conclude that it is unlikely that Trumpler 37 is indeed 
physically related to Cep OB2.
Our age estimate of $3-6$ Myr encompasses determinations by other 
authors 
(e.g.~Clayton \& Fitzpatrick 1987\nocite{clayton87};
Marschall et al.~1990\nocite{marschall90}).


\end{document}